\documentclass[aps, 11pt, preprint, prc, nofootinbib, superscriptaddress]{revtex4-1}

\usepackage[usenames]{color}
\usepackage{graphicx}
\usepackage{amsmath}
\usepackage{amsfonts}
\usepackage{amssymb}
\usepackage{mathrsfs}
\usepackage{bm}
\usepackage{verbatim}
\usepackage{cancel}
\usepackage{CJK}

\newcommand{\mhi}{M_{\text{hi}}}
\newcommand{\MNN}{M_{NN}}
\newcommand{\chn}[3]{{{}^{#1}\!{#2}_{#3}}}
\newcommand{\cs}[2]{\chn{#1}{S}{#2}}
\newcommand{\cp}[2]{\chn{#1}{P}{#2}}
\newcommand{\cd}[2]{\chn{#1}{D}{#2}}
\newcommand{\cf}[2]{\chn{#1}{F}{#2}}
\newcommand{\cg}[2]{{}^{#1}{G}_{#2}}
\newcommand{\csd}{{\cs{3}{1}-\cd{3}{1}}}
\newcommand{\cpf}{{\cp{3}{2}-\cf{3}{2}}}
\newcommand{\cdg}{{\cd{3}{3}-\cg{3}{3}}}

\graphicspath{{./plots/}}

\preprint{CTP-SCU/2018002}

\begin{document}

\title{Perturbative $NN$ scattering in chiral effective field theory}

\author{Shaowei Wu}

\author{Bingwei Long}
\email{bingwei@scu.edu.cn}
\affiliation{College of Physical Science and Technology, Sichuan University, Chengdu, Sichuan 610065, China}

\date{February 23, 2019}

\begin{abstract}
Within the framework of chiral effective field theory, perturbative calculation for $NN$ scattering is carried out in partial waves with orbital angular momentum $L \geqslant 1$. The primary goal is to identify the lowest angular momenta at which perturbative treatment of chiral forces can apply. Results up to the order where the subleading two-pion exchange appears are shown. It is concluded that perturbation theory applies to all partial waves but $\cs{1}{0}$, $\csd$, and $\cp{3}{0}$. Where it is applicable, perturbation theory with the delta-less chiral forces produces good agreement with the empirical phase shifts up to $k_\text{c.m.} \simeq 300$ MeV.
\end{abstract}

\maketitle

\section{Introduction\label{sec:intro}}

Low-energy nuclear theory is challenging because of the complicated, nonperturbative structure of atomic nuclei. On the other hand, the construction of nuclear forces with chiral effective field theory (EFT) involves so many terms of $NN$ and/or higher body contributions that appear in the form of Feynman diagrams~\cite{Weinberg90, Weinberg91, Ordonez-1993tn, Ordonez-1995rz, Epelbaum-1999dj, Entem-2002sf, Ekstrom-2013kea, Epelbaum-2014sza, Entem-2014msa, Gezerlis-2014zia}. It could be rewarding to identify a limited number of the most crucial pieces of nuclear forces so that the complexity of nuclear theory would be reduced. One-pion exchange (OPE) has long been thought as the most important long-range nuclear force, but it gets weakened by the centrifugal barrier as the orbital angular momentum $L$ increases. Our goal in the present paper is to investigate the lowest angular momenta at which OPE can be accounted for as perturbation, as opposed to iteration-to-all-order treatment in the $S$ waves. Two-pion exchanges (TPEs) and contact interactions will be considered as well, as parts of a systematic chiral EFT framework for perturbative $NN$ scattering.

An extreme case of similar efforts was made two decades ago by Kaplan, Savage, and Wise (KSW)~\cite{Kaplan-1998tg, Kaplan-1998we, Beane-2008bt}. In the KSW scheme, OPE is considered as perturbative even in the $S$ waves, for momenta softer than $\MNN \equiv 16\pi f_\pi^2/(g_A^2 m_N) \simeq 290$ MeV. $\MNN$  characterizes the strength of OPE and it is considerably smaller than $\mhi \sim 1$ GeV, the momentum scale below which chiral EFT applies. KSW hoped that the nonperturbative feature of nuclear forces is completely attributed to a pair of two-body and one three-nucleon contact terms, exactly like in pionless EFT~\cite{bira-proceeding, Kaplan-1998tg, Kaplan-1998we, vanKolck-1998bw, Bedaque-1998kg, Bedaque-1999ve}. This idea is attractive because the KSW scheme would then offer an opportunity to build nuclear physics around the unitarity limit~\cite{Konig-2016utl}, with pion exchanges and higher-order contact terms treated as the perturbations that displace real-world nuclear phenomena away from the ideal unitarity limit. Unfortunately, it turned out to be overly optimistic to take $\MNN$ as the breakdown scale of perturbative OPE; an actual calculation~\cite{Fleming-1999ee} showed that validity window of the KSW scheme does not seem to be considerably larger than that of pionless EFT, which has a still simpler structure.

Nonperturbative iteration of OPE is, however, complicated, not only for computational complexity but for interfering the counting of contact interactions. When OPE is weak enough to justify strict perturbation theory, renormalization does not typically surprise us as to estimating the size of contact terms: They will obey naive dimensional analysis (NDA)~\cite{Manohar-1983md, Weinberg90, Weinberg91, Georgi93}. But when OPE is so strong as to necessitate resummation to all orders, the sizes of contact terms may be dramatically different than given by NDA. As shown in Ref.~\cite{Nogga-2005hy}, in the partial waves where OPE is singular and attractive, a counterterm must be at leading order (LO) in order for the scattering amplitude to satisfy renormalization-group (RG) invariance, i.e., to absorb ultraviolet (UV) cutoff dependence induced by nonperturbative iteration of OPE. For partial waves with orbital angular momentum $L > 0$, like $\cp{3}{0}$, $\cpf$, $\cd{3}{2}$, etc., this means that contact terms, even though they are momentum dependent, must be promoted several powers relative to their NDA ordering. Also on the grounds of RG invariance, Ref.~\cite{Birse-2005um} used different technique to reach a similar conclusion that some of the counterterms need to be promoted, although the detail of power counting is different from that of Ref.~\cite{Nogga-2005hy}. References~\cite{Valderrama-2009ei, Valderrama-2011mv, Long-2007vp, Long-2011xw, Long-2012ve, Long-2013cya} studied separately renormalization of TPEs on top of nonperturbative OPE. They both found that promotion of counterterms is propagated to higher orders: The more attractive triplet channels included at LO, the more counterterms than NDA would have assigned must be considered. For instance, following Ref.~\cite{Long-2011xw}, one will have two counterterms for $\cd{3}{2}$ up to next-to-next-to-next-to-leading order (N$^3$LO) \footnote{We follow the convention of, for example, Refs.~\cite{PavonValderrama-2016lqn, SanchezSanchez-2017tws} to label the orders. Next-to-leading order (NLO) is the order relatively smaller than LO by $\mathcal{O}(Q/\mhi)$, which was considered vanishing in Weinberg's power counting scheme. So our N$^3$LO corresponds to next-to-next-to-leading order (N$^2$LO) in, for example, Ref.~\cite{Machleidt-2011zz}.}, whereas NDA assigns none. (For works touching upon issues of renormalization on chiral forces but from different viewpoints, see Refs.~\cite{Yang-2004ss, Yang-2007hb, Entem-2007jg, Yang-2009kx, Epelbaum-2009sd, Szpigel-2011bc, Marji-2013uia, Ren-2016jna, Epelbaum-2017byx, Ren-2017yvw}.)

Although this proliferation of short-range parameters is not necessarily a serious setback, thanks to the wealth of few-nucleon data, the complication of power counting due to nonperturbative OPE creates motivation to identify possible perturbative components of OPE, because that would cause chiral nuclear forces become more amenable to simple analysis like NDA. Another obvious mechanism to weaken OPE, besides softer momenta, is the centrifugal barrier. As a matter of fact, it is well known from empirical phase-shift analyses~\cite{Stoks-1993tb, Arndt-2007qn, SAID} that partial waves with $L \simeq 3$ contribute little to $NN$ scattering; therefore, OPE must become weak enough at certain $L$ to warrant perturbation theory, even for momenta $Q \gtrsim m_\pi$.

Reference~\cite{Birse-2005um} had a pioneering study on the correlation between angular momenta and perturbativeness of the tensor part of OPE. The solution to the Schr\"odinger equation as a complex function of $p$/$\MNN$ was examined for its analyticity, where $p$ is the center-of-mass (c.m.) momentum. For each partial wave, there exists a critical value $p_c$ for the solution to remain meromorphic in $p$/$\MNN$. The branch cut above $p_c$/$\MNN$ would invalidate the Taylor expansion in $p$/$\MNN$, or equivalently, perturbative treatment of OPE. Reproduced in Table~\ref{tab:birse} is the critical value $p_c$ for each of the lowest few of triplet channels. However, contact interactions could also be a function of $\MNN$ because renormalization mixes the short-range part of OPE and contact terms~\cite{Beane-2000wh}, and this dependence was not accounted for in the analysis of Ref.~\cite{Birse-2005um}. The missing information on the role played by contact interactions can only be supplemented by the underlying theory or $NN$ data. In summary, our attitude towards Table~\ref{tab:birse} is that while it provides a guideline, the value of $p_c$ listed there does not have to be the clear-cut breakdown point for perturbative OPE, so the empirical $NN$ phase shifts will be indispensable in our study.

\begin{table}
  \caption{According to Ref.~\cite{Birse-2005um}, critical values of the CM momentum for triplet channels above which OPE is nonperturbative.}
  \label{tab:birse}
  \centering
  \begin{tabular}{cc}
  \hline
  \hline
  Channel & $p_c$ (MeV) \\
  \hline
  $\csd$ & $66$ \\
  $\cp{3}{0}$ & $182$ \\
  $\cp{3}{1}$ & $365$ \\
  $\cpf$ & $470$ \\
  $\cd{3}{2}$ & $403$ \\
  $\cdg$ & $382$ \\
  \hline
  \hline
  \end{tabular}
\end{table}

Reference~\cite{PavonValderrama-2016lqn} examined systematically suppression factors for OPE in the spin-singlet channels where OPE is regular, as opposed to having singular asymptote $\to r^{-3}$ when $r \to 0$. The emphasis was on the ``$L$ counting'' that tells one not only whether OPE is perturbative for a given $L$ but which order to place it in the counting hierarchy. For instance, while OPE in $\cp{1}{1}$ is counted as NLO, its projection onto $\cf{1}{3}$ is counted as N$^2$LO. This is especially useful in some many-body calculations where inclusion of fewer relative partial waves costs less computing resource.

We will cover both singlet and triplet channels in the present paper, but it is not our goal to understand the impact of $L$ on power counting as rigorously as in Ref.~\cite{PavonValderrama-2016lqn}. We wish to know the lowest angular momentum where OPE is already perturbative for momenta relevant for most nuclear structure calculations. In this exploratory work, OPE will be simply counted as NLO for all the partial waves studied here. We lean on empirical phase shift values to decide whether perturbative treatment is applicable. Because TPEs have more singular short-range behavior $r^{-5}$ or $r^{-6}$, it is even more difficult to quantify their centrifugal suppression \textit{a priori}. So we assume the most natural case in which TPEs are suppressed by the same ratio as OPE, i.e., the relative order of $Q^2/\mhi^2$ is assumed unchanged for any partial waves.

The paper is structured as follows. We explain in Sec.~\ref{sec:pc} the perturbative power counting used for calculations. The results are shown in Sec.~\ref{sec:results} where discussions are offered. Finally, we summarize in Sec.~\ref{sec:discussion}.

\section{Power counting\label{sec:pc}}

A brief summary of chiral EFT in $NN$ scattering will be helpful. Those who are interested in more comprehensive reviews can consult, for instance, Refs.~\cite{Bedaque-2002mn, Epelbaum-2008ga, Machleidt-2011zz}. OPE is the usual starting point,
\begin{equation}
  V_{1\pi}(\vec{q}\,) = - \frac{g_A^2}{4f_\pi^2} \bm{\tau}_1 \bm{\cdot} \bm{\tau}_2  \frac{\vec{\sigma}_1 \cdot \vec{q}\, \vec{\sigma}_2 \cdot \vec{q}}{\vec{q}\,^2 + m_\pi^2}  \, , \label{eqn:OPE}
\end{equation}
where the pion decay constant $f_\pi = 92.4$ MeV, axial coupling $g_A = 1.29$, isospin-averaged pion mass $m_\pi = 138$ MeV, and $\vec{q}$ is the momentum transfer between nucleon $1$ and $2$. The once iteration of OPE is given by
\begin{equation}
 V_{1\pi} G_0 V_{1\pi} = \int \frac{d^3 l}{(2\pi)^3} V^\Lambda_{1\pi}(\vec{k} - \vec{l}) \frac{m_N}{k^2 - l^2 + i\epsilon} V^\Lambda_{1\pi}(\vec{l} - \vec{k}')
  \, , \label{eqn:iterOPE}
\end{equation}
where $\vec{k}$ ($\vec{k}'$) is the incoming (outgoing) momentum in the c.m. frame and $G_0$ is the free propagator. Iterations with $G_0$ are regularized with a momentum cutoff regulator of separable Gaussian form:
\begin{equation}
  V^{\Lambda}(\vec{p}\,', \vec{p}\,) \equiv \exp\left(-\frac{\vec{p}\,'^4}{\Lambda^4}\right) V(\vec{p}\,', \vec{p}\,) \exp\left(-\frac{\vec{p}\,^4}{\Lambda^4}\right) \, . \label{eqn:Gaussian}
\end{equation}

Ignoring for a moment suppression by the centrifugal barrier and considering $Q \sim m_\pi$, where $Q$ denotes typical size of external momenta, we can invoke standard ChPT power counting~\cite{Manohar-1983md, Weinberg90, Weinberg91, Georgi93} and estimate OPE as
\begin{equation}
V_{1\pi} \sim \frac{4\pi}{m_N} \frac{1}{\MNN}\, . \label{eqn:countingOPE}
\end{equation}
The loop integral in $V_\pi G_0 V_\pi$ has well-known enhancement proportional to $m_N$, with a typical numerical factor different than relativistic loop integrals:
\begin{equation}
  \int \frac{d^3 l}{(2\pi)^3} \frac{m_N}{k^2 - l^2 + i\epsilon} \sim \frac{m_N Q}{4\pi} \, .
\end{equation}
With these elements, the once-iterated OPE is counted as
\begin{equation}
V_{1\pi} G_0 V_{1 \pi} \sim \left(\frac{4\pi}{m_N} \frac{1}{\MNN}\right)^2 \frac{m_N Q}{4\pi} \sim \frac{4\pi}{m_N} \frac{Q}{\MNN^2} \, . \label{eqn:countingVGV}
\end{equation}
The moral of this qualitative analysis is that one can expect a kinematic window of small momenta $Q \ll \MNN$, in which OPE can be treated in perturbation theory~\cite{Kaplan-1998tg, Kaplan-1998we}. But $\MNN$ is not a mass scale tied to a definite observable, for example, mass of a particle, so the definition of $\MNN$ as the OPE strength is inevitably murky. This means the usefulness of expansion in $Q/\MNN$ will crucially depend on the numerical factors floating around in Eq.~\eqref{eqn:iterOPE}. A series of higher order calculations~\cite{Fleming-1999ee, Soto-2007pg} showed that the KSW scheme is not as promising as many had hoped, despite the effort to remedy it~\cite{ Beane-2008bt}.

\subsection{Centrifugal suppression of OPE}

The perturbative scheme used in the present paper does not require $Q \ll \MNN$. We instead use the fact that orbital angular momentum $L$ can also suppress long-range forces like pion exchanges. Centrifugal suppression of $V_{1\pi}$ and its once iteration can be expressed symbolically as
\begin{equation}
  \langle L' | V_{1\pi} | L \rangle \sim \frac{4\pi}{m_N} \frac{1}{a(L)\, \MNN} \, , \quad \langle L' |  V_{1\pi} G_0 V_{1 \pi} | L \rangle \sim \frac{4\pi}{m_N} \frac{1}{a(L)\, \MNN} \frac{Q}{b(L) \MNN}\, ,\label{eqn:albl}
\end{equation}
where $L' = L$ or $L+2$. $a(L)$ describes how the Born approximation of OPE is suppressed after partial-wave projection. $b(L)$ serves as a notice that $V_{1\pi}$ and its iterations are not necessarily suppressed identically. This is evidenced by the analytical expressions of partial-wave amplitudes of $V_{1\pi}$ and $V_{1\pi} G_0 V_{1 \pi}$ shown in Eq. (39) of Ref.~\cite{Fleming-1999ee}. Although the chiral limit was used there, those expressions suffice to make the point that numerical factor $a(L)$ and $b(L)$ are in principle different.

A better understanding of $a(L)$ and $b(L)$ is more relevant for larger $L$ if one is interested in how much exactly OPE is weakened for a given $L$, a goal similar to that of Ref.~\cite{PavonValderrama-2016lqn}. We concern ourselves, however, in the present paper with a different task that emphasizes identifying the critical value of $L$ where OPE starts to be perturbative for $Q \sim m_\pi$, rather than on quantifying the suppression, as done in Ref.~\cite{PavonValderrama-2016lqn}. So we take a more simplistic point of view towards $a(L)$ and $b(L)$ that assumes $a(L) \simeq b(L)$ and $a(L) \MNN \sim \mhi$. We will apply this counting to partial waves considered in this paper: $1 \leqslant L \leqslant 4$.

When $a(L) \leqslant 1$,  OPE is nonperturbative and requires iterations to all orders. We know at least that this is the case for both $S$ waves. These partial-wave amplitudes are LO. When $a(L)$ is sufficiently large so that $a(L)\MNN \sim \mhi$, OPE will be perturbative enough to be placed at NLO and the LO amplitude for this partial wave vanishes. So the tree-level OPE, its once and twice iterations are NLO, N$^2$LO, N$^3$LO, and so on, respectively,
\begin{equation}
  \begin{split}
  \text{NLO} :& \qquad T^{(1)}_{1\pi} = V_{1\pi} \sim \frac{4\pi}{m_N} \frac{1}{\mhi} \sim \frac{4\pi}{m_N} \frac{1}{\MNN} \frac{Q}{\mhi} \, , \\
  \text{N$^2$LO}:& \qquad T^{(2)}_{1\pi} = V_{1\pi} G_0 V_{1\pi} \sim \frac{4\pi}{m_N} \frac{1}{\MNN} \frac{Q^2}{\mhi^2} \, , \\
  \text{N$^3$LO}:& \qquad T^{(3)}_{1\pi} = V_{1\pi} G_0 V_{1\pi} G_0 V_{1\pi} \sim \frac{4\pi}{m_N} \frac{1}{\MNN} \frac{Q^3}{\mhi^3} \, ,
  \end{split}
\end{equation}
where we have used $Q \sim m_\pi \sim \MNN$.

\subsection{Leading and subleading TPEs\label{sec:TPEs}}

The leading TPE, denoted by $V_{2\pi}^{(0)}$, is made up of one-loop irreducible diagrams with $\nu = 0$ vertexes~\cite{Ordonez-1993tn, Friar-1994zz, Ordonez-1995rz, Kaiser-1997mw}, where $\nu$ is the chiral index defined by Weinberg~\cite{Weinberg90, Weinberg91}. If the centrifugal barrier is ignored, the absence of pure $NN$ intermediate states makes it straightforward to count $V_{2\pi}^{(0)}$:
\begin{equation}
V_{2\pi}^{(0)} \sim \frac{1}{f_\pi^2} \frac{Q^2}{\mhi^2} \sim \frac{4\pi}{m_N} \frac{1}{\MNN} \frac{Q^2}{\mhi^2} \, .
\end{equation}
The subleading TPE $V_{2\pi}^{(1)}$ has $\nu = 1$ $\pi\pi NN$ ``seagull'' couplings, thus one order higher than $V_{2\pi}^{(0)}$. $V_{2\pi}^{(1)}$ received much attention because the uncertainty of these $\nu = 1$ $\pi\pi NN$ couplings contribute significantly to theoretical errors of chiral nuclear forces.

We use the expressions for TPEs found in Ref.~\cite{Kaiser-1997mw}. In order to help define normalization convention, we reproduce $V_{2\pi}^{(0)}$ here as follows:
\begin{equation}
  V_{2\pi}^{(0)}(\vec{q}\,) = \bm{\tau}_1 \bm{\cdot} \bm{\tau}_2 W_C(q) + \vec{\sigma}_1 \cdot \vec{\sigma}_2 V_S(q)  +\vec{\sigma}_1 \cdot \vec{q}\, \vec{\sigma}_2 \cdot \vec{q}\, V_T(q) \, , \nonumber
\end{equation}
where
\begin{equation}
\begin{split}
  W_C(q) &= - \frac{1}{384 \pi^2 f_\pi^4} \left[4m_\pi^2\left(5g_A^4 - 4g_A^2 - 1\right) + q^2 \left(23g_A^4 - 10g_A^2 - 1\right) + \frac{48g_A^4 m_\pi^4}{4m_\pi^2 + q^2} \right] L(q)\, , \\
  V_T(q) &= -\frac{V_S(q)}{q^2} = - \frac{3 g_A^4}{64\pi^2 f_\pi^4} L(q) \, , \label{eqn:TPE0WV}
\end{split}
\end{equation}
with
\begin{equation}
  L(q) = \frac{w}{q} \ln \frac{w + q}{2m_\pi}\, , \quad w = \sqrt{4m_\pi^2 + q^2}\, .
\end{equation}
Note that the sign convention for potentials is different than that of Ref.~\cite{Kaiser-1997mw}. Terms that are polynomials in $q^2$ or $m_\pi^2$ have been dropped because they can be absorbed into contact terms. We have gone one step further and have also dropped terms proportional to $q^2 \ln m_\pi$ that have impacts on chiral extrapolation of lattice QCD results.

Now we turn to centrifugal suppression of long-range parts of TPEs. Regardless of the value of $m_\pi$, $V_{2\pi}^{(0)}$ ($V_{2\pi}^{(1)}$) has singularity $r^{-5}$ ($r^{-6}$) for $r \to 0$. With such singularities, it is difficult to quantify centrifugal suppression of TPEs to the level of sophistication of analyses in Refs.\cite{Birse-2005um, PavonValderrama-2016lqn}. However, for our purpose of examining the lowest partial waves where perturbation theory is at all valid, it suffices to explore the simplest scenario in which TPEs are assumed to be suppressed by the same power as is OPE, i.e., the relative difference between TPEs and OPE remains $(Q/\mhi)^2$ for any $L > 0$: $V_{2\pi}^{(0)}$ ($V_{2\pi}^{(1)}$) is N$^3$LO (N$^4$LO).

We collect below the pion-exchange amplitudes, up to N$^4$LO where the subleading TPE starts to contribute:
\begin{align}
  T_{\pi}^{(2)} &= V_{1\pi} G_0 V_{1\pi} \, , \label{eqn:TpiN2LO_SUP} \\
  T_{\pi}^{(3)} &= V_{1\pi} \left(G_0 V_{1\pi}\right)^2 + V_{2\pi}^{(0)} \, ,  \label{eqn:TpiN3LO_SUP} \\
  T_{\pi}^{(4)} &= V_{1\pi} \left(G_0 V_{1\pi}\right)^3 + \left(V_{1\pi} G_0 V_{2\pi}^{(0)} + \text{perm.} \right) + V_{2\pi}^{(1)}  \label{eqn:TpiN4LO_SUP} \, ,
\end{align}
where ``perm.'' refers to all possible permutations of potentials appearing in the iteration.

\subsection{Contact interactions}

Since the one-loop diagrams for TPEs are divergent, we need counterterms to render the loop integrals finite. Called primordial counterterms ~\cite{Long-2011xw}, they are second-degree polynomials in momenta, and hence for $L \geqslant 1$ they act only on $P$ waves. By NDA, these $Q^2$ counterterms are counted as N$^2$LO. Since TPEs are counted as N$^3$LO or higher, it is tempting to push these second-degree polynomials to higher orders as well. However, aside from high-momentum modes of TPEs, other short-range physics can drive these counterterms too, and they may be less weakened by the centrifugal barrier than TPEs. To be prudent, we count contact terms at least as assigned by NDA. Therefore, the second-degree-polynomial counterterms are always at N$^2$LO:
\begin{equation}
\langle \text{chn} \, , p' | V_\text{ct}^{(2)} | \text{chn} \, , p \rangle = C_\text{chn}^{(0)} p' p \, ,
  \label{eqn:VSNNLO}
\end{equation}
where chn = $\cp{1}{1}$, $\cp{3}{0}$, $\cp{3}{1}$, and $\cp{3}{2}$. Iterations involving both $V_\text{ct}^{(2)}$ and $V_{1\pi}$ contribute to higher order $P$-wave amplitudes:
\begin{align}
  T_{\pi, \text{ct2}}^{(3)} &= V_\text{ct}^{(2)} G_0 V_{1\pi} + \text{perm.} \, , \label{eqn:piCT2_3}\\
  T_{\pi, \text{ct2}}^{(4)} &= V_\text{ct}^{(2)} G_0 V_\text{ct}^{(2)} + \left[ V_\text{ct}^{(2)} \left(G_0 V_{1\pi}\right)^2 + \text{perm.} \right] \, . \label{eqn:piCT2_4}
\end{align}

At N$^3$LO, one needs to renormalize loop integrals appearing on the right-hand side of Eqs.~\eqref{eqn:TpiN3LO_SUP} and \eqref{eqn:piCT2_3}. The UV divergence can be superficially estimated and the counterterm to remove the divergence has the form of second-degree polynomials in momenta but now with different coefficient than that of $V_\text{ct}^{(2)}$:
\begin{equation}
  \langle \text{chn} \, , p' | V_\text{ct}^{(3)} | \text{chn} \, , p \rangle = C_\text{chn}^{(1)} p' p \, , \label{eqn:Vct3_SUP}
\end{equation}
where, again, chn = $\cp{1}{1}$, $\cp{3}{0}$, $\cp{3}{1}$, and $\cp{3}{2}$.

At N$^4$LO, one needs the following counterterms to remove the divergences on the right-hand side of Eqs.\eqref{eqn:TpiN4LO_SUP} and \eqref{eqn:piCT2_4}. For uncoupled $P$ waves,
\begin{equation}
\langle \text{chn}\, , p' | V_\text{ct}^{(4)} | \text{chn}\, , p \rangle = \left[C_\text{chn}^{(2)} + D_\text{chn}^{(0)} \left(p^2 + p'^2\right) \right] p' p \, ,
 \label{eqn:VCTN4LOPWave_SUP}
\end{equation}
where chn $= \cp{1}{1}$, $\cp{3}{0}$, and $\cp{3}{1}$.
For most $D$ waves,
\begin{equation}
\langle \text{chn}\, , p' | V_\text{ct}^{(4)} | \text{chn}\, , p \rangle = D_\text{chn}^{(0)} {p'}^2 p^2 \, , \label{eqn:VCTN4LODWave_SUP}
\end{equation}
where chn $= \cd{1}{2}$, $\cd{3}{2}$, $\cd{3}{3}$.
Finally, for coupled channel $\cp{3}{2} - \cf{3}{2}$,
\begin{equation}
  \langle \cpf \, , p' | V_\text{ct}^{(4)} | \cpf \, , p \rangle = p' p
  \begin{pmatrix}
C_\cp{3}{2}^{(2)} + D^{(0)}_\cp{3}{2}({p'}^2 + p^2) & E^{(0)}_{PF}\, p^2 \\
E^{(0)}_{PF}\, {p'}^2 & 0
\end{pmatrix} \, .\\
\end{equation}

We are now in the position to assemble the partial-wave amplitudes:
\begin{align}
T^{(2)} &=  V_\text{ct}^{(2)} + T^{(2)}_\pi\, , \\
T^{(3)} &=  T_{\pi, \text{ct2}}^{(3)} + T^{(3)}_\pi + V_\text{ct}^{(3)} \, , \\
T^{(4)} &=  T_{\pi, \text{ct2}}^{(4)} + T^{(4)}_\pi + \left(V_\text{ct}^{(3)} G_0 V_{1\pi} + \text{perm.} \right) + V_\text{ct}^{(4)} \, .
\end{align}

\section{Results and Discussions\label{sec:results}}

When converting expansion of the scattering amplitude $T$ to phase shifts and mixing angles, one needs to respect the unitarity of the $S$ matrix according to power counting so that at any given order breaking of the unitarity is always in higher order. An example of how this is done can be found in the appendix of Ref.~\cite{Long-2011xw}.

We have not yet explained one important ingredient about the chiral forces that have been laid out in the previous section: The couplings of $\nu = 1$ $\pi \pi NN$ seagull vertexes. Called $c_i$'s, they decide crucially the size of subleading TPE, $V_{2\pi}^{(1)}$. We use the values of $c_i$'s extracted from an analysis of $\pi N$ scattering data that was based on the Roy-Steiner equation~\cite{Hoferichter-2015tha, Hoferichter-2015hva}. These values are listed in Table~\ref{tab:cis}, where the orders refer to ChPT expansions of the $\pi N$ scattering amplitude. The $\pi N$ amplitude related to Table~\ref{tab:cis} does not have the explicit degrees of freedom of the delta isobar, which is compatible with the delta-less TPE's we have adopted in the paper.

\setlength{\tabcolsep}{+9pt}
\begin{table}
  \caption{The central values of $c_i$'s used in the paper, in unit of GeV$^{-1}$. They are extracted from an analysis based on the Roy-Steiner equation of $\pi N$ scattering data~\cite{Hoferichter-2015tha, Hoferichter-2015hva}. The orders refer to ChPT counting of the $\pi N$ scattering amplitude.}
  \label{tab:cis}
  \centering
  \begin{tabular}{lccc}
  \hline
  \hline
  \qquad & NLO \qquad & N$^2$LO \qquad & N$^3$LO \\
  \hline
  $c_1$ & -0.74 \qquad & -1.07 \qquad & -1.10 \qquad \\
  $c_3$ & -3.61 & -5.32 & -5.54 \\
  $c_4$ & 2.44 & 3.56 & 4.17 \\
  \hline
  \hline
  \end{tabular}
\end{table}
\setlength{\tabcolsep}{-9pt}

We will use in this paper the rather large uncertainty of $c_i$'s to our advantage to probe the role of the delta isobar in $NN$ scattering. Before doing that, however, we use the ``NLO'' set of Table~\ref{tab:cis} to study how the perturbative formulation fares in each partial wave. Only towards the end of this section will the uncertainty of $c_i$'s be investigated.

\begin{figure}
  \centering
  \includegraphics[scale=0.31]{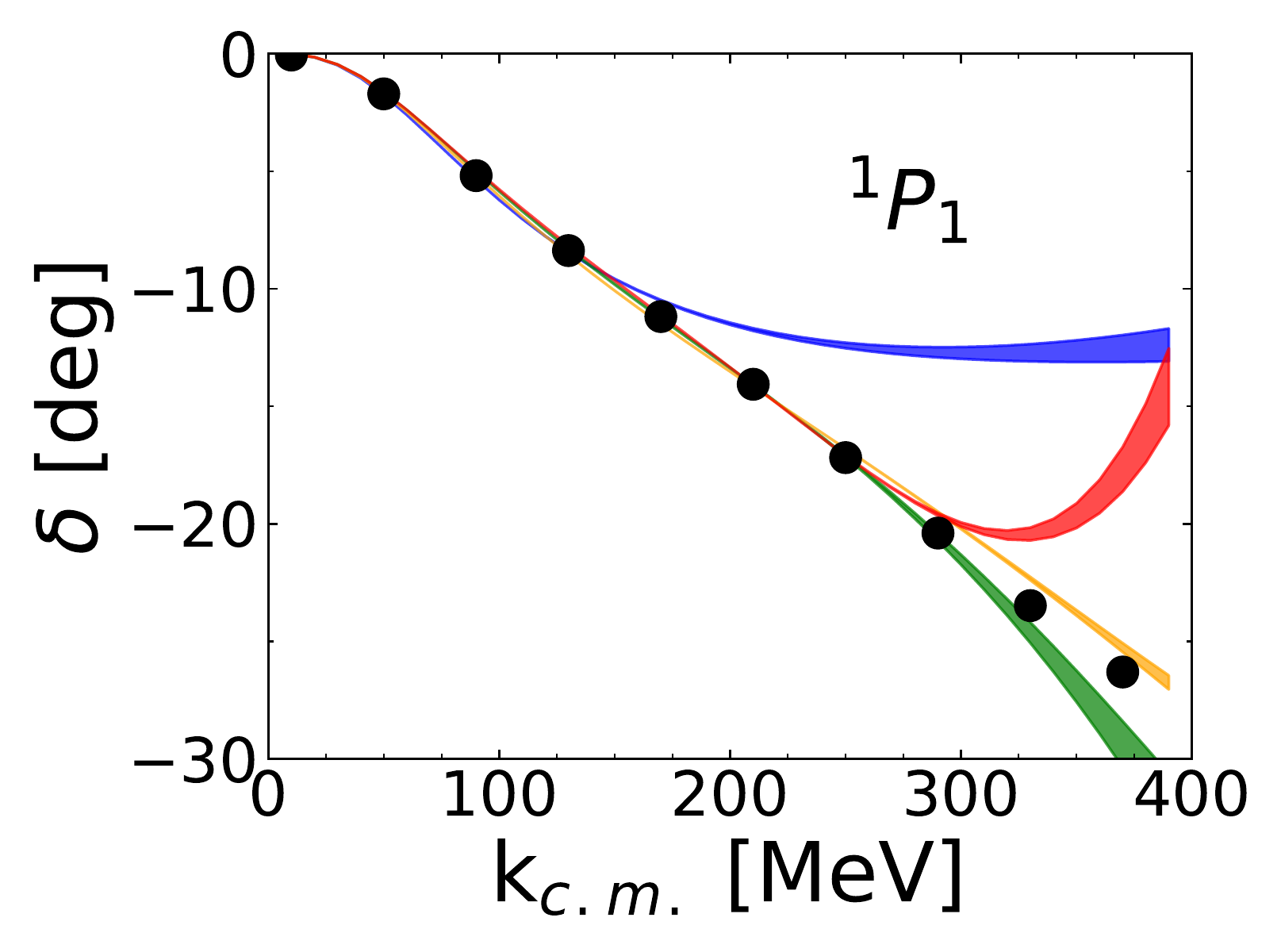}
  \includegraphics[scale=0.31]{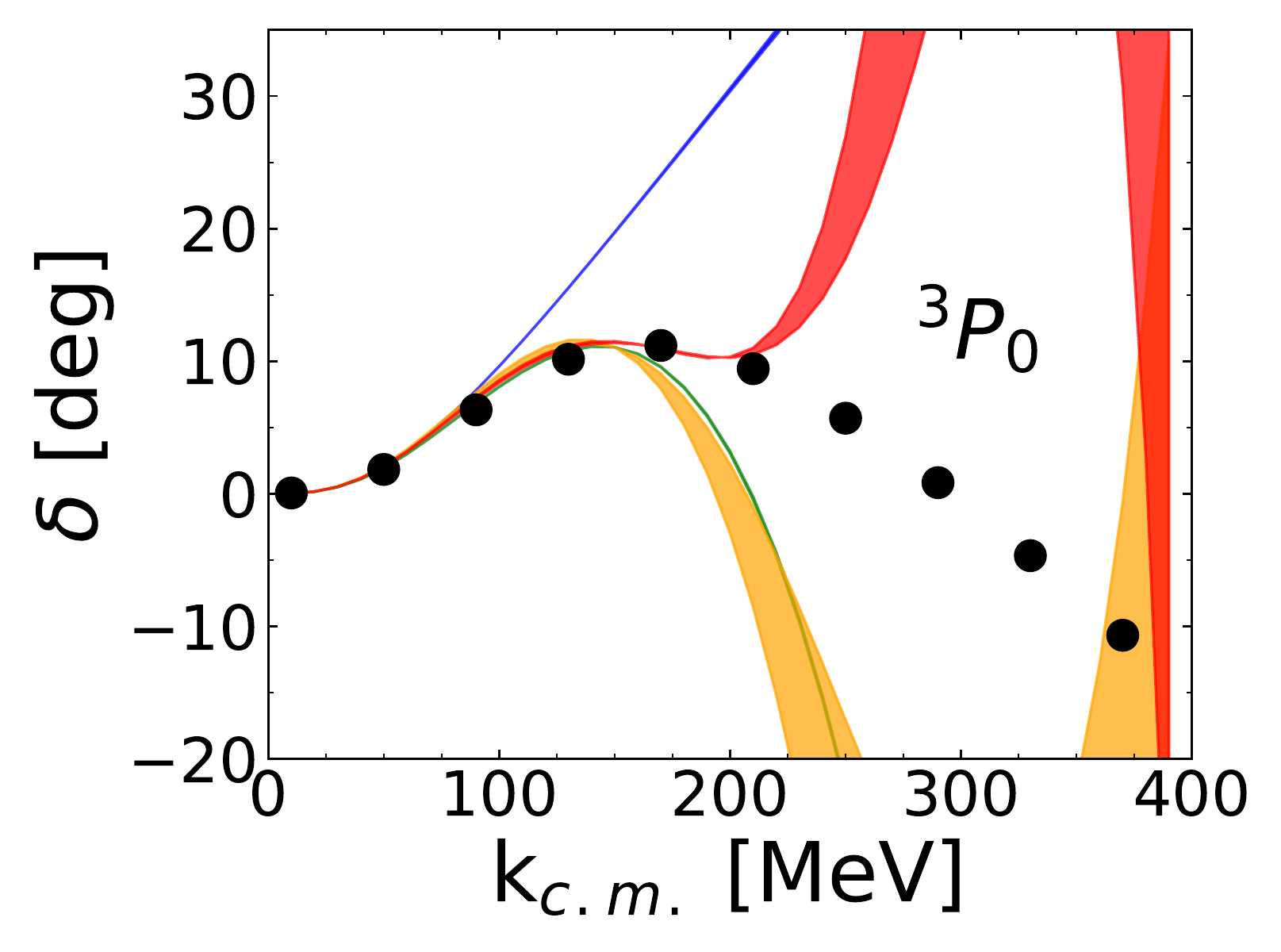}
  \includegraphics[scale=0.31]{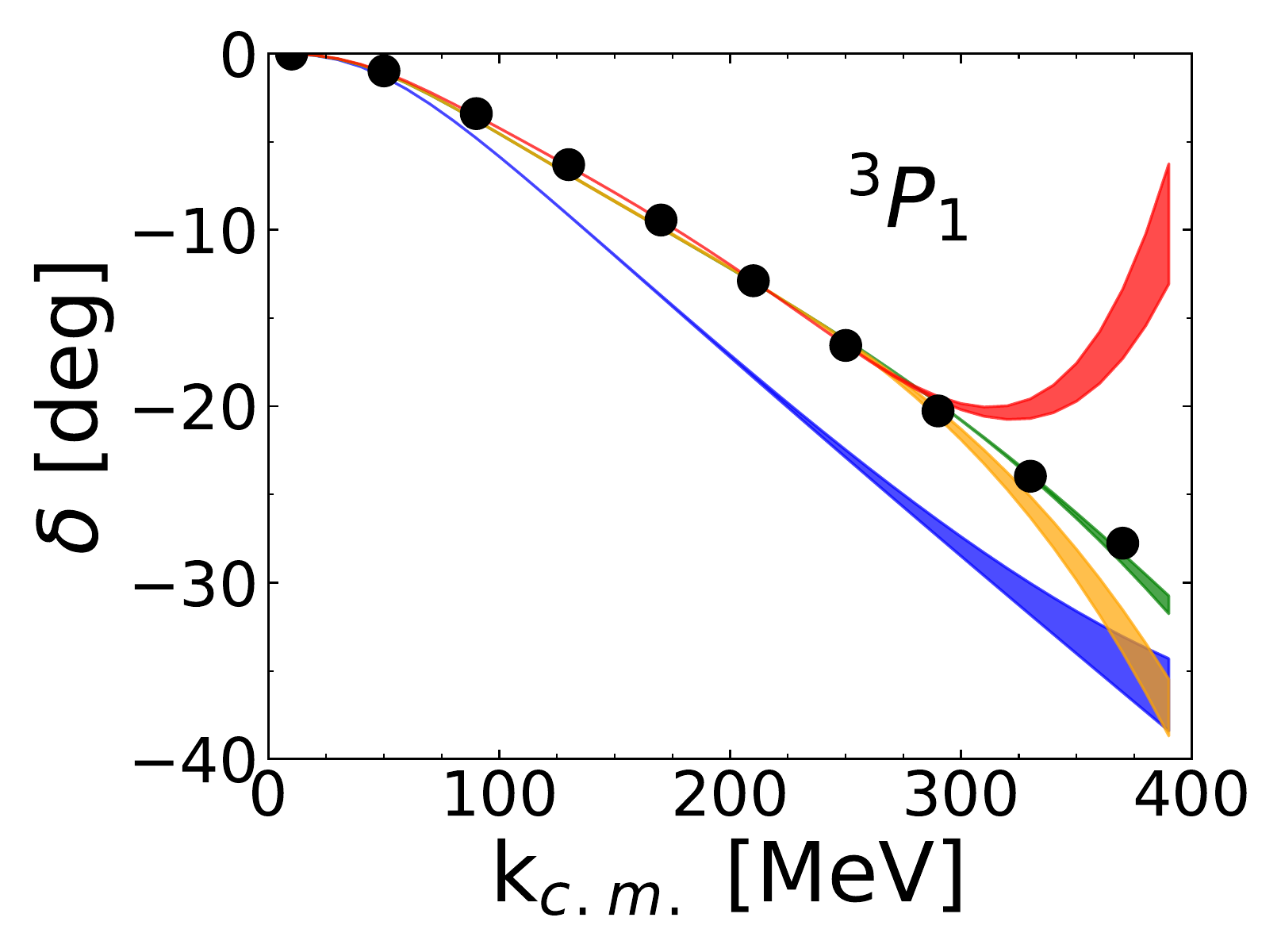}
  \caption{$P$-wave $NN$ phase shifts as functions of the c.m. momentum. The solid circles are the empirical phase shifts from the SAID program~\cite{SAID}. The blue, green, orange and red bands correspond respectively to NLO, N$^2$LO, N$^3$LO and N$^4$LO. The bands are generated for a range of cutoff values. See the text for more explanations.}
  \label{fig:PwaveBands}
\end{figure}

Shown in Fig.~\ref{fig:PwaveBands} are the results for the $P$ waves except for $\cp{3}{2}$. In fitting to the empirical phase shifts provided by the SAID program at the George Washington University (GWU)~\cite{Arndt-2007qn, SAID}, we have generally favored points near $k_\text{c.m.} = m_\pi$, between $k_\text{c.m.} = 130$ and $200$ MeV. The cutoff value $\Lambda$ is varied from 0.8 to 4.8 GeV, except for $\cp{3}{0}$, which uses $\Lambda = 0.8 - 2.4$ GeV.

In Fig.~\ref{fig:PwaveBands}, $\cp{3}{0}$ stands out not only for its failure at N$^4$LO to describe the phase shifts beyond $k_\text{c.m.} \simeq 200$ MeV, but also for the failure to converge: The order-to-order change blows up rapidly above $k_\text{c.m.}
\simeq 200$ MeV, except for the change from N$^2$LO to N$^3$LO. This is very much in accord with the critical momentum obtained in Ref.~\cite{Birse-2005um} for $\cp{3}{0}$, $p_c = 182$ MeV (see Table~\ref{tab:birse}).

N$^4$LO involves a three-loop integral, as expressed in Eq.~\eqref{eqn:TpiN4LO_SUP}. This is especially troublesome for $\cp{3}{0}$ in our numerical calculation because the singular attraction of OPE in $\cp{3}{0}$ is significantly stronger than in other $P$ waves, resulting in more divergent integrals. The great sensitivity to the cutoff value makes it more difficult to obtain the amplitude when subtracting two large numbers numerically. This turns out to prevent us from going beyond $\Lambda = 2.4$ GeV for $\cp{3}{0}$ at N$^4$LO. So we adopt $\Lambda = 0.8 - 2.4$ GeV for all orders in $\cp{3}{0}$. In addition, N$^3$LO and N$^4$LO have large cutoff variation beyond $k_\text{c.m.} \simeq 200$ MeV even for $\Lambda < $2.4 GeV, and they are not (nor need to be) fully shown in the current scope of the plot.

\begin{figure}
  \centering
  \includegraphics[scale=0.33]{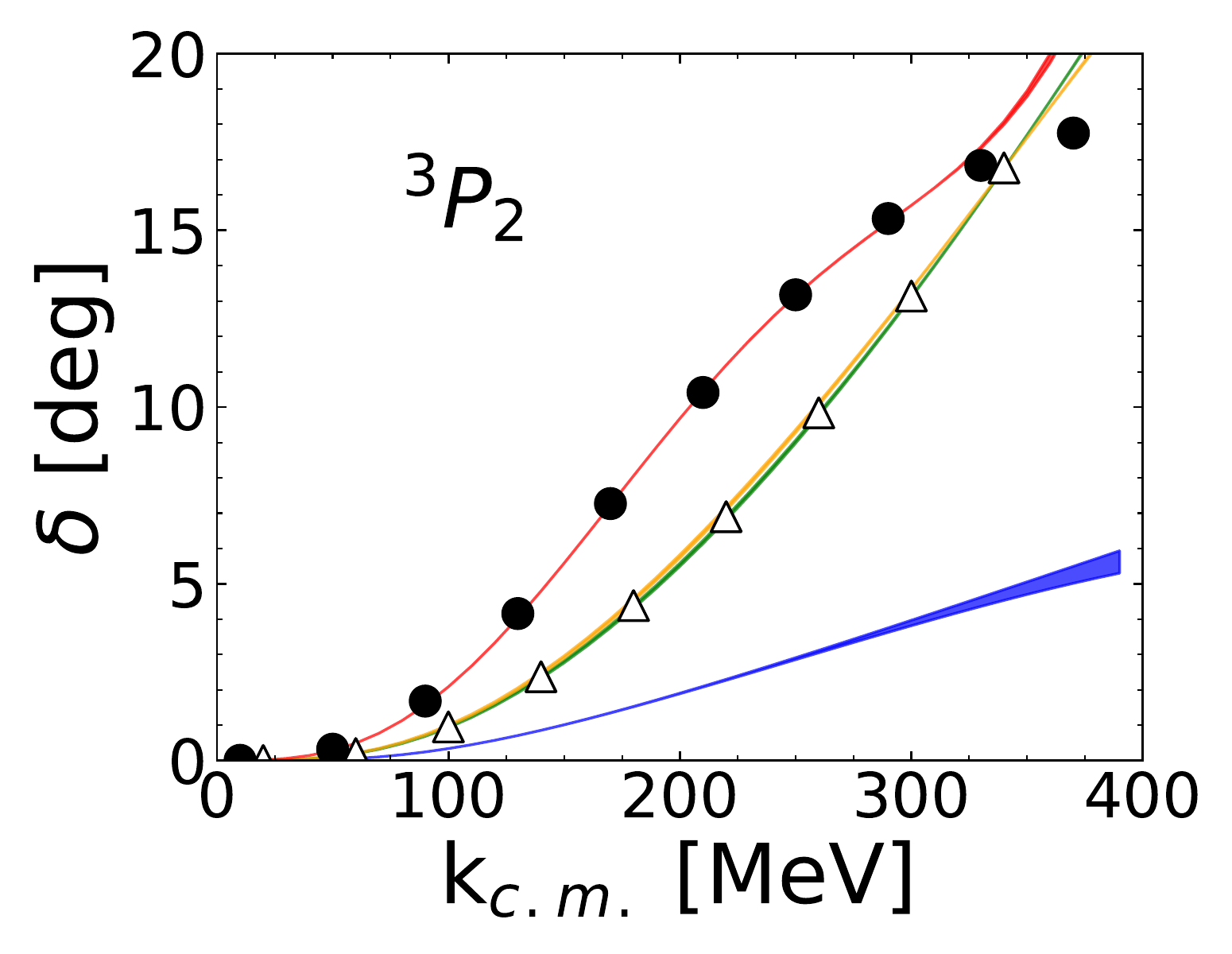}
  \includegraphics[scale=0.33]{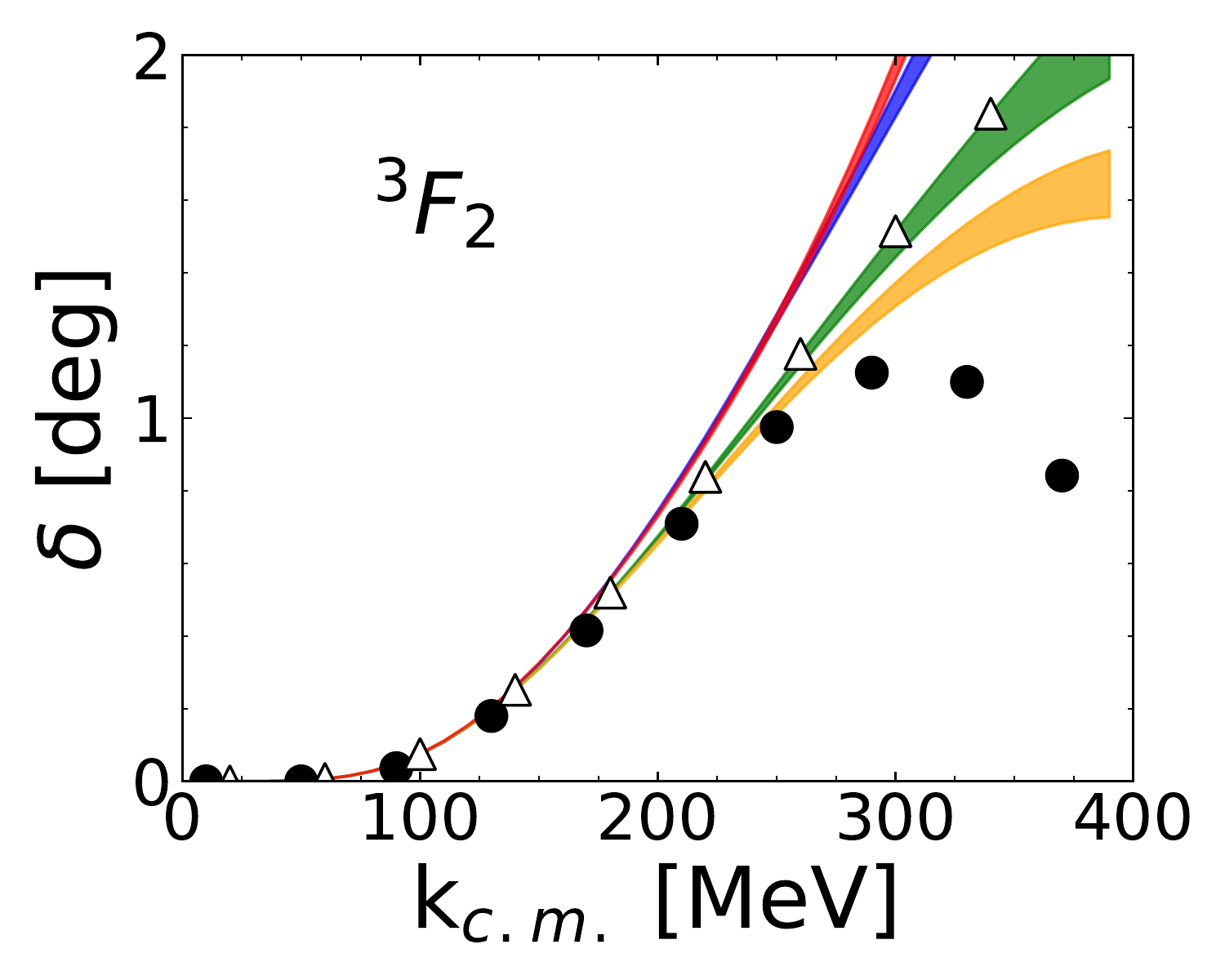}
  \includegraphics[scale=0.33]{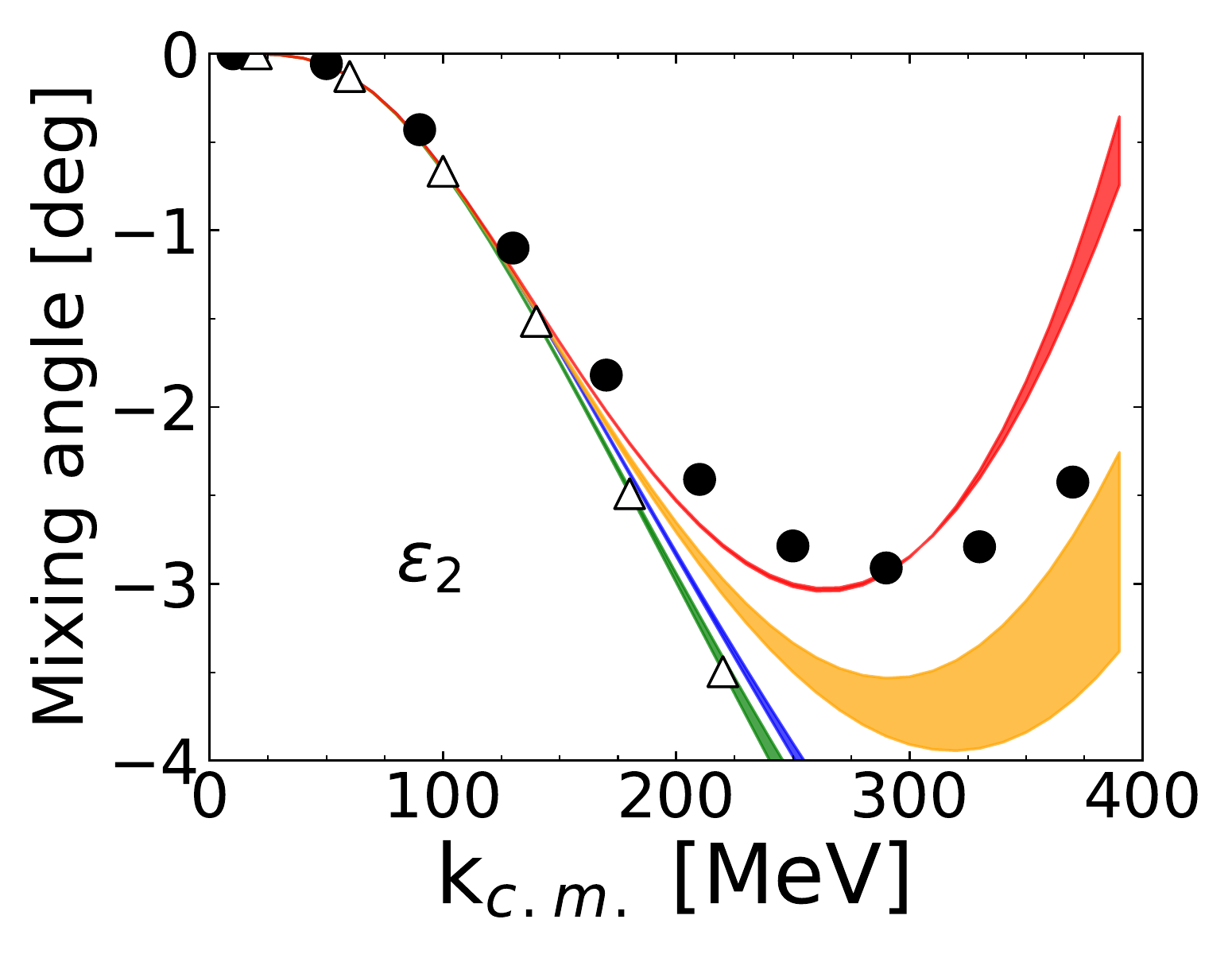}
  \caption{$\cpf$ phase shifts and mixing angle. The triangles represent the sum of OPE and its once iteration ($V_{1\pi} + V_{1\pi} G_0 V_{1\pi}$) for $\Lambda \to \infty$. For explanation of other symbols, see Fig.~\ref{fig:PwaveBands}.}
  \label{fig:3P2Bands}
\end{figure}

Another channel calling for special attention is $\cpf$, which is shown in Fig.~\ref{fig:3P2Bands}. We first remark that the NLO contribution to $\cp{3}{2}$ is notably weaker than other $P$ waves. This is more quantitatively reflected by $P$-wave scattering volumes at NLO:
\begin{equation}
a_\cp{1}{1} = 2 \MNN^{-1} m_\pi^{-2}\, , \quad a_\cp{3}{0} = -2 \MNN^{-1} m_\pi^{-2}\, , \quad a_\cp{3}{1} = -\frac{4}{3} \MNN^{-1} m_\pi^{-2}\, , \quad a_\cp{3}{2} = 0 \, , \label{eqn:Pwavevols}
\end{equation}
which can be computed straightforwardly by applying $k/m_\pi \to 0$ to the expressions for partial-wave projections of OPE found in Ref.~\cite{Fleming-1999ee}. We see that while the OPE contribution to other $P$ waves is in line with expectation based on dimensional analysis, it vanishes in $\cp{3}{2}$.

Due to surprisingly weak strength of OPE in $\cp{3}{2}$, the $\cp{3}{2}$ contact terms dominate the phase shifts at N$^{2}$LO and N$^{3}$LO. But we do not want to tune the $\cp{3}{2}$ contact coupling to such large values that it destroys the convergence. On the other hand, we notice that the sum of the tree-level OPE and the once-iterated OPE is finite: $V_{1\pi} + V_{1\pi}G_0 V_{1\pi}$, which are depicted in Fig.~\ref{fig:PwaveBands} by triangles. They can be evaluated numerically for large enough $\Lambda$ or using the KSW ``NNLO'' expression worked out in Ref.~\cite{Fleming-1999ee} with dimensional regularization. We take these triangles around $k = m_\pi$ in $\cp{3}{2}$ as somewhat ``natural'' values for N$^{2, 3}$LO to be fitted to. Then, at N$^{4}$LO, we switch back to fitting to the empirical phase shifts. In summary, because the Born approximation of OPE is accidentally weak in $\cp{3}{2}$, some additional care to $\cpf$ in fitting procedure could be taken in order to improve order-by-order convergence.

\begin{figure}[tb]
  \centering
  \includegraphics[scale=0.33]{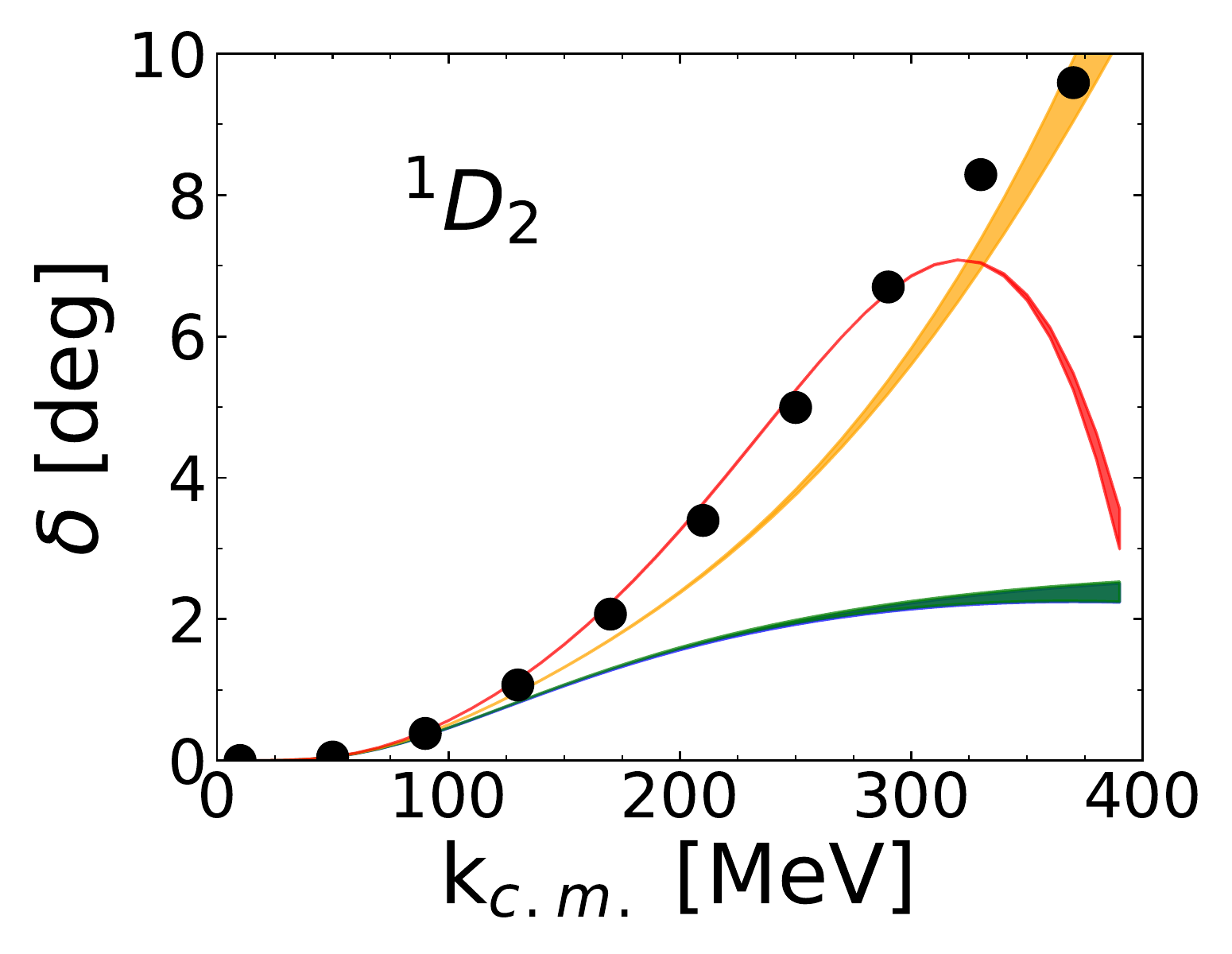}
  \includegraphics[scale=0.33]{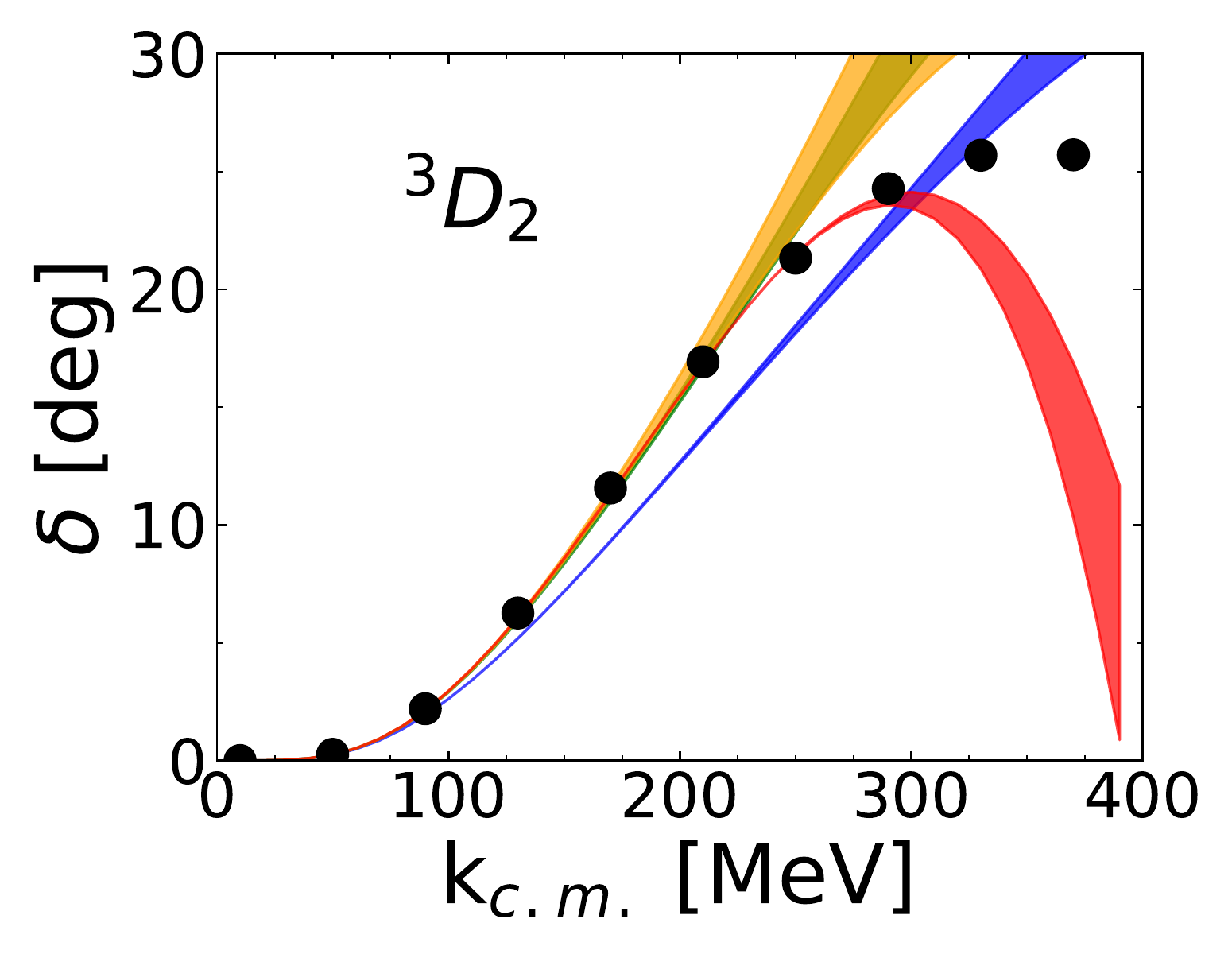} \\
  \includegraphics[scale=0.33]{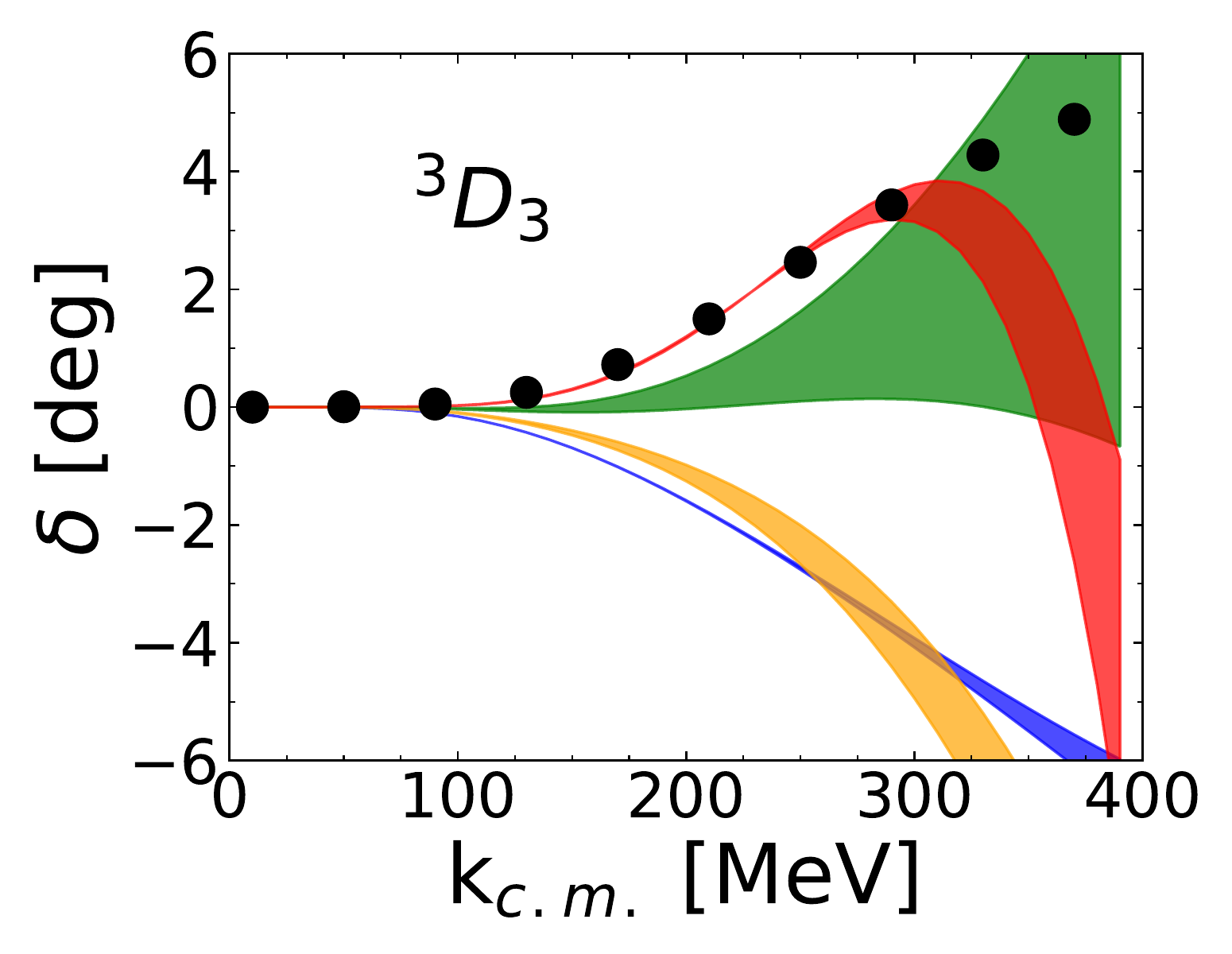}
  \includegraphics[scale=0.33]{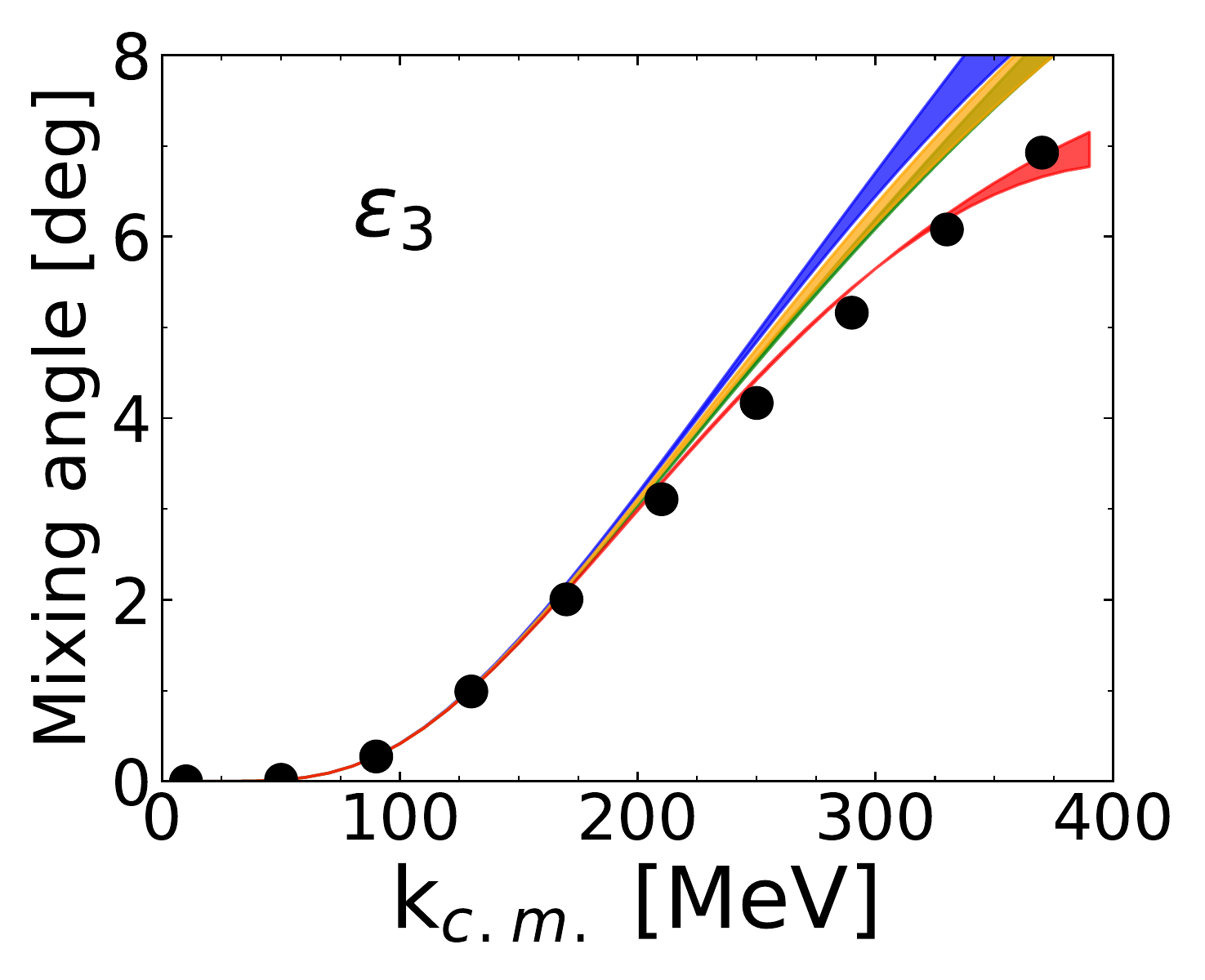}
   \caption{$D$-wave phase shifts and mixing angel $\mathcal{E}_3$. For explanation of symbols, see Fig.~\ref{fig:PwaveBands}.}
  \label{fig:DwaveBands}
\end{figure}

\begin{figure}[tb]
  \centering
  \includegraphics[scale=0.33]{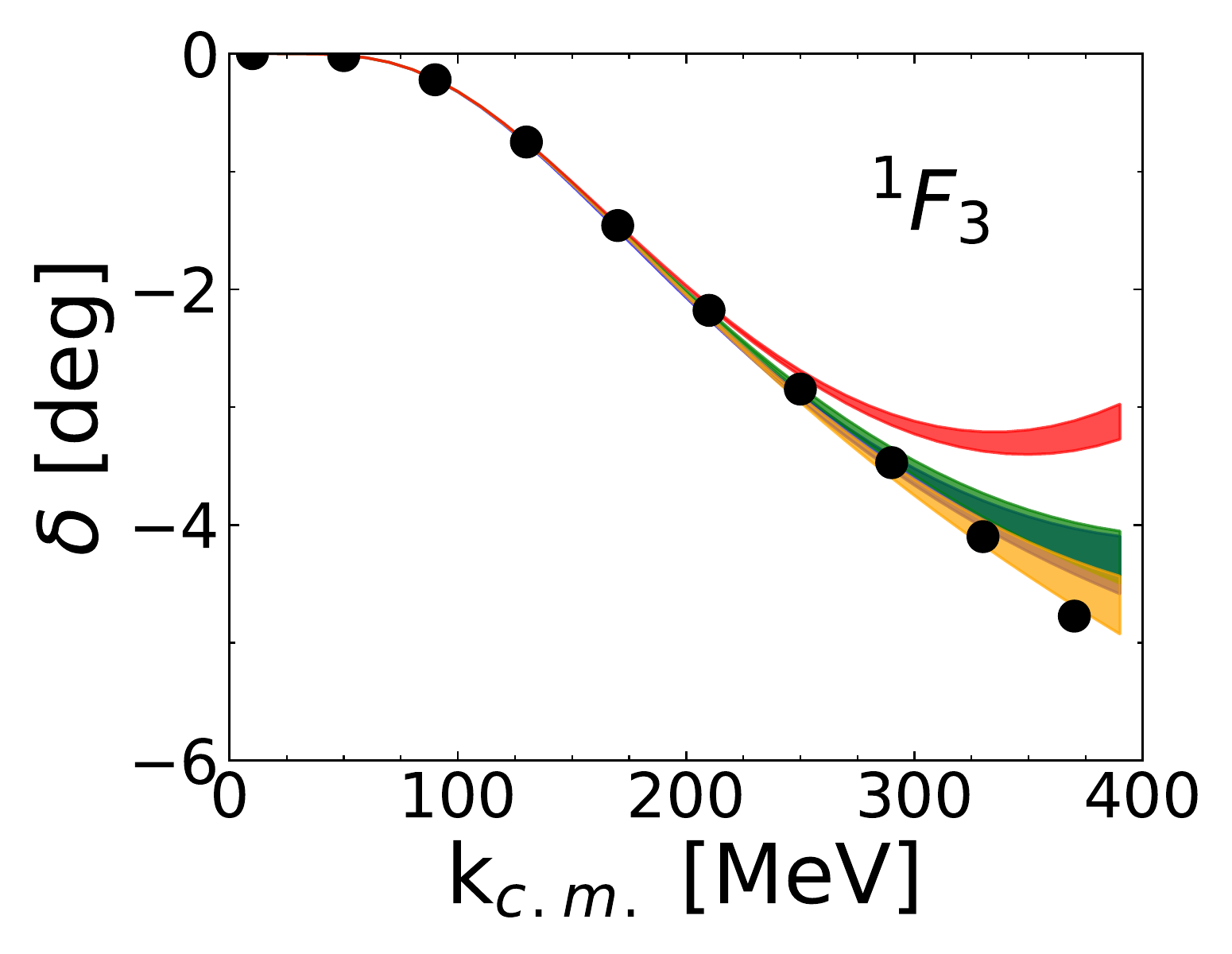}
  \includegraphics[scale=0.33]{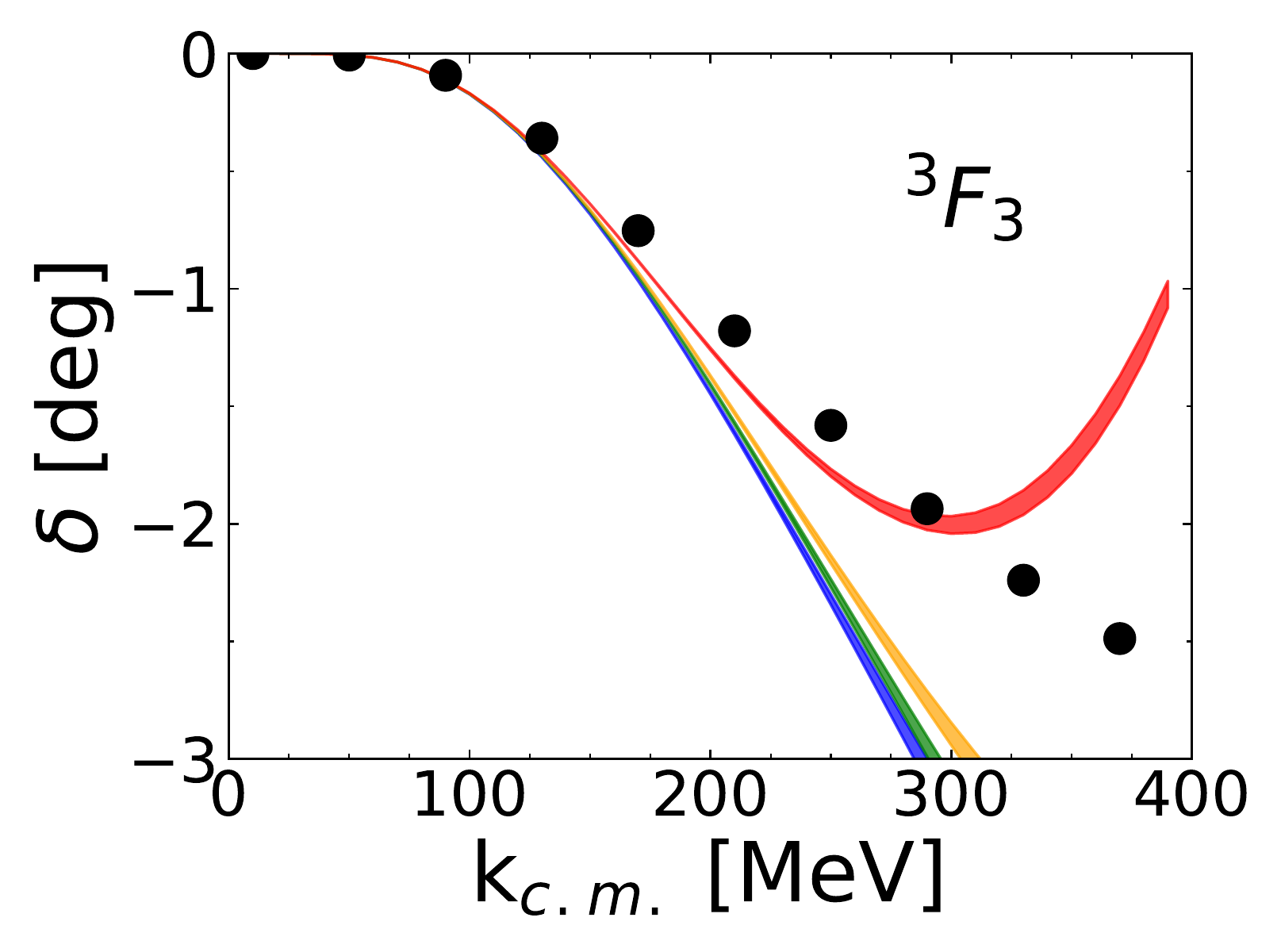}
  \includegraphics[scale=0.33]{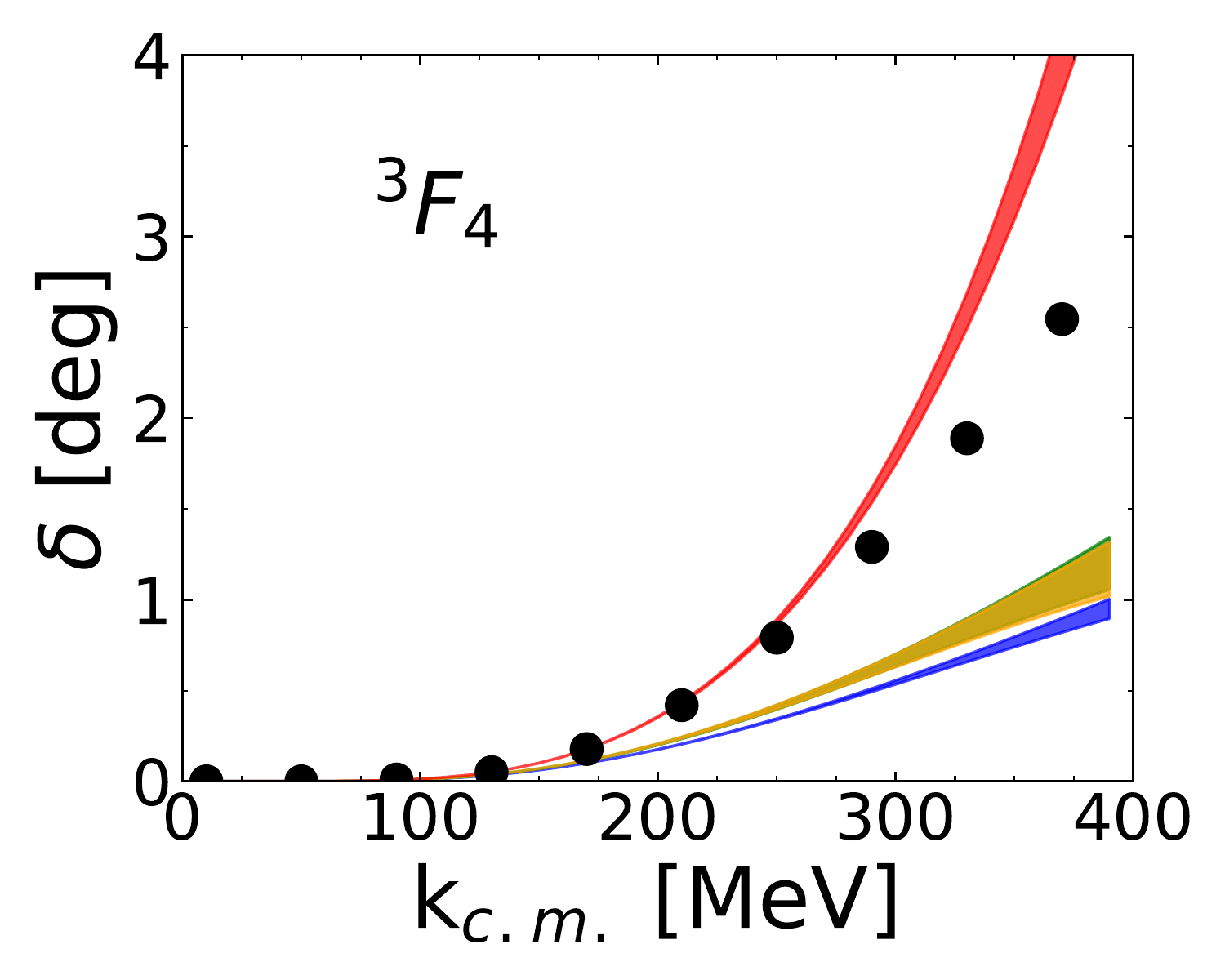}
  \caption{$F$-wave phase shifts. For explanation of symbols, see Fig.~\ref{fig:PwaveBands}.}
  \label{fig:FwaveBands}
\end{figure}

\begin{figure}[tb]
  \centering
  \includegraphics[scale=0.33]{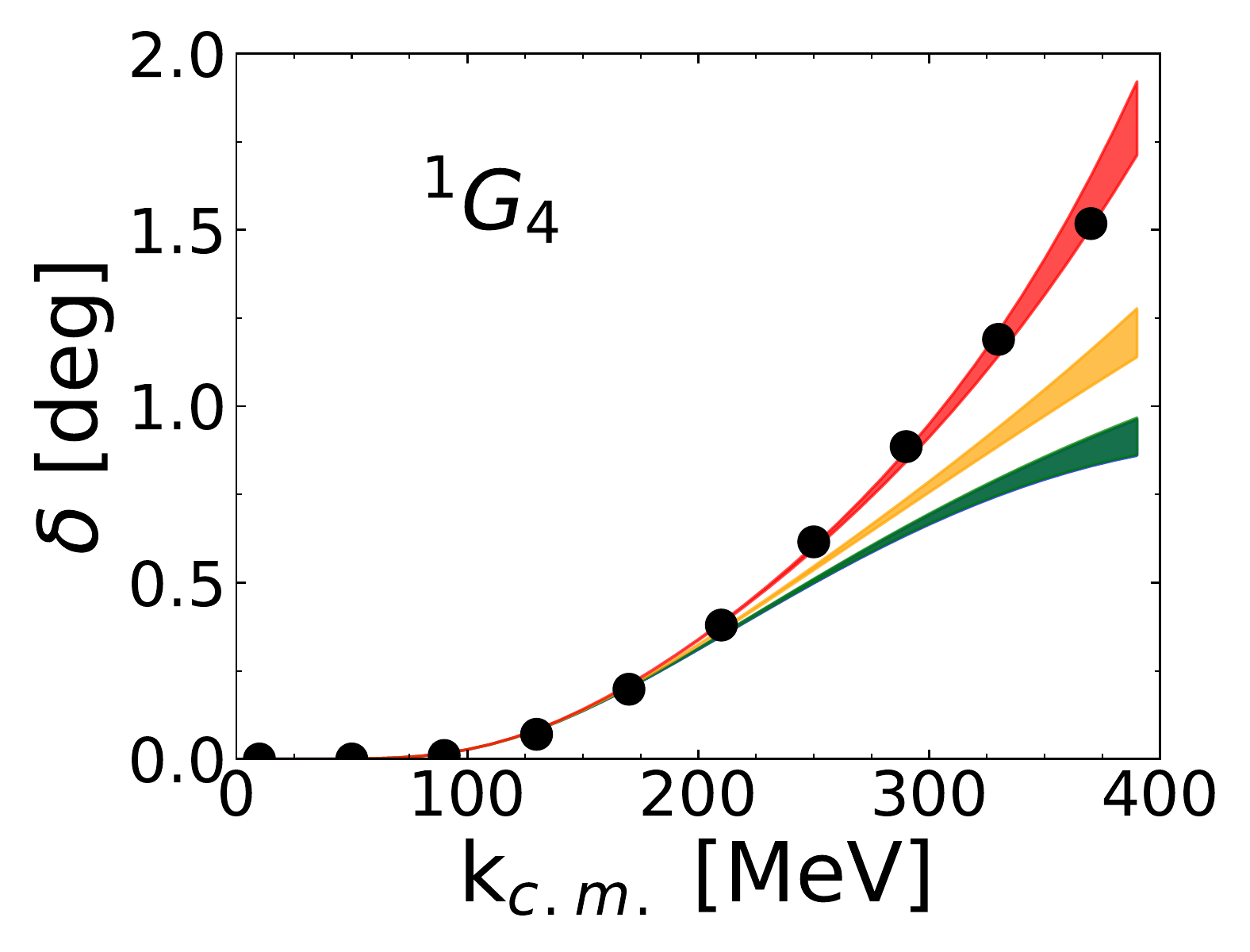}
  \includegraphics[scale=0.33]{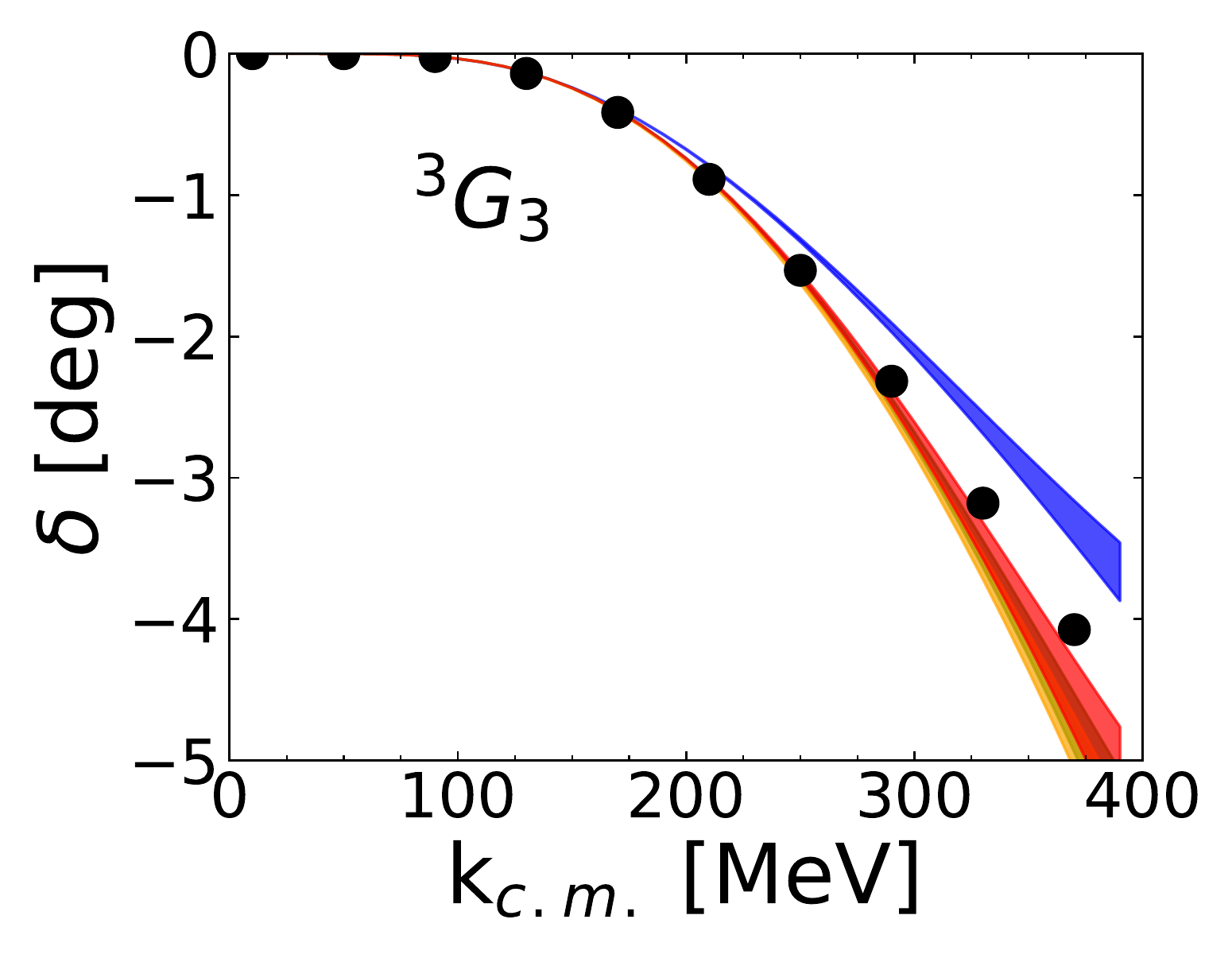} \\
  \includegraphics[scale=0.33]{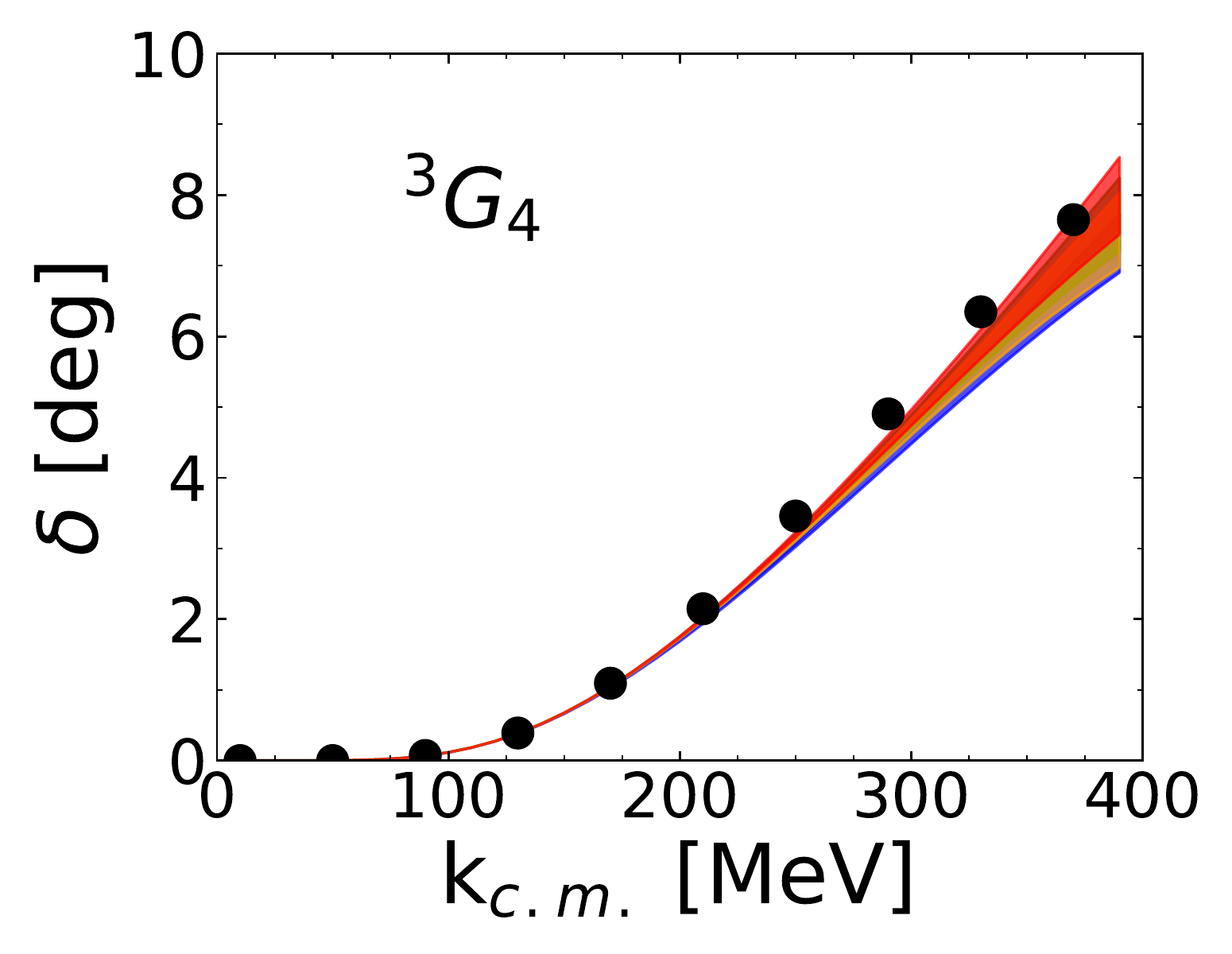}
  \includegraphics[scale=0.33]{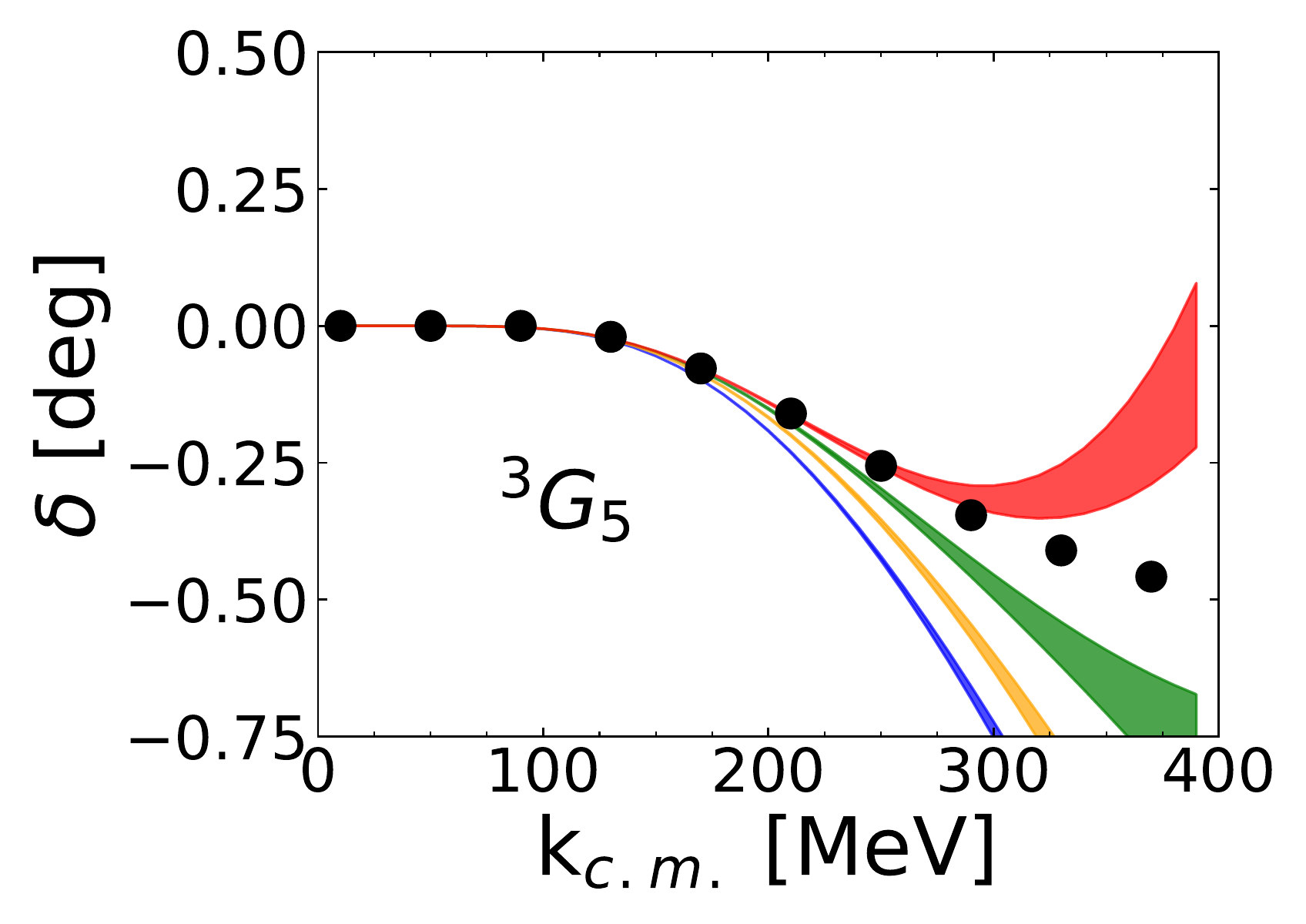}
  \caption{$G$-wave phase shifts. For explanation of symbols, see Fig.~\ref{fig:PwaveBands}.}
  \label{fig:GwaveBands}
\end{figure}

Let us now apply the perturbative formulation to $D$, $F$, $G$ waves and mixing angles $\mathcal{E}_3$. The results are shown in Figs.~\ref{fig:DwaveBands} - \ref{fig:GwaveBands}. The bands are generated by cutoff values from $\Lambda = 0.8$ to $4.8$ GeV. Because $\cd{3}{1}$ is coupled to $\cs{3}{1}$ through the rather strong tensor force of OPE, it is considered here as a nonperturbative channel mostly for convenience, also in accord with Ref.~\cite{Birse-2005um}.

With increasing orders, the EFT amplitudes generally show systematically improved agreement with the empirical phase shifts. N$^4$LO differs from the GWU phase shifts by about $1$ deg in all channels at $k_\text{c.m.} \simeq 300$ MeV. The EFT expansion appears to break down beyond $k_\text{c.m.} \simeq 300$ MeV, which is especially the case for $P$, $D$, $F$ waves, and mixing angle $\mathcal{E}_2$. This is, however, expected from any delta-less framework.

As stated at the beginning of this section, the large uncertainty of $c_i$'s in Table~\ref{tab:cis} can be exploited to probe the role of the delta isobar in $NN$ scattering, which is integrated out in the delta-less theory used here. Due to the relative smallness of nucleon-delta mass splitting $\delta \simeq 300$ MeV, the ChPT expansion of the $\pi N$ scattering amplitude converges slowly; therefore, the EFT truncation error, due mostly to lack of the delta-isobar degrees of freedom, dominates the uncertainty of $c_i$'s. When folded into the $NN$ scattering amplitude for N$^4$LO, this uncertainty of $c_i$'s is expected to dictate theoretical errors of the $NN$ amplitude at momenta where the delta resonance is ``felt'' by $NN$ scattering data. In other words, the variation due to different sets of $c_i$'s will indicate the momenta where the delta isobar can no longer be viewed as short-range physics that can be absorbed into $NN$ contact terms.

\begin{figure}
  \centering
  \includegraphics[scale=0.30]{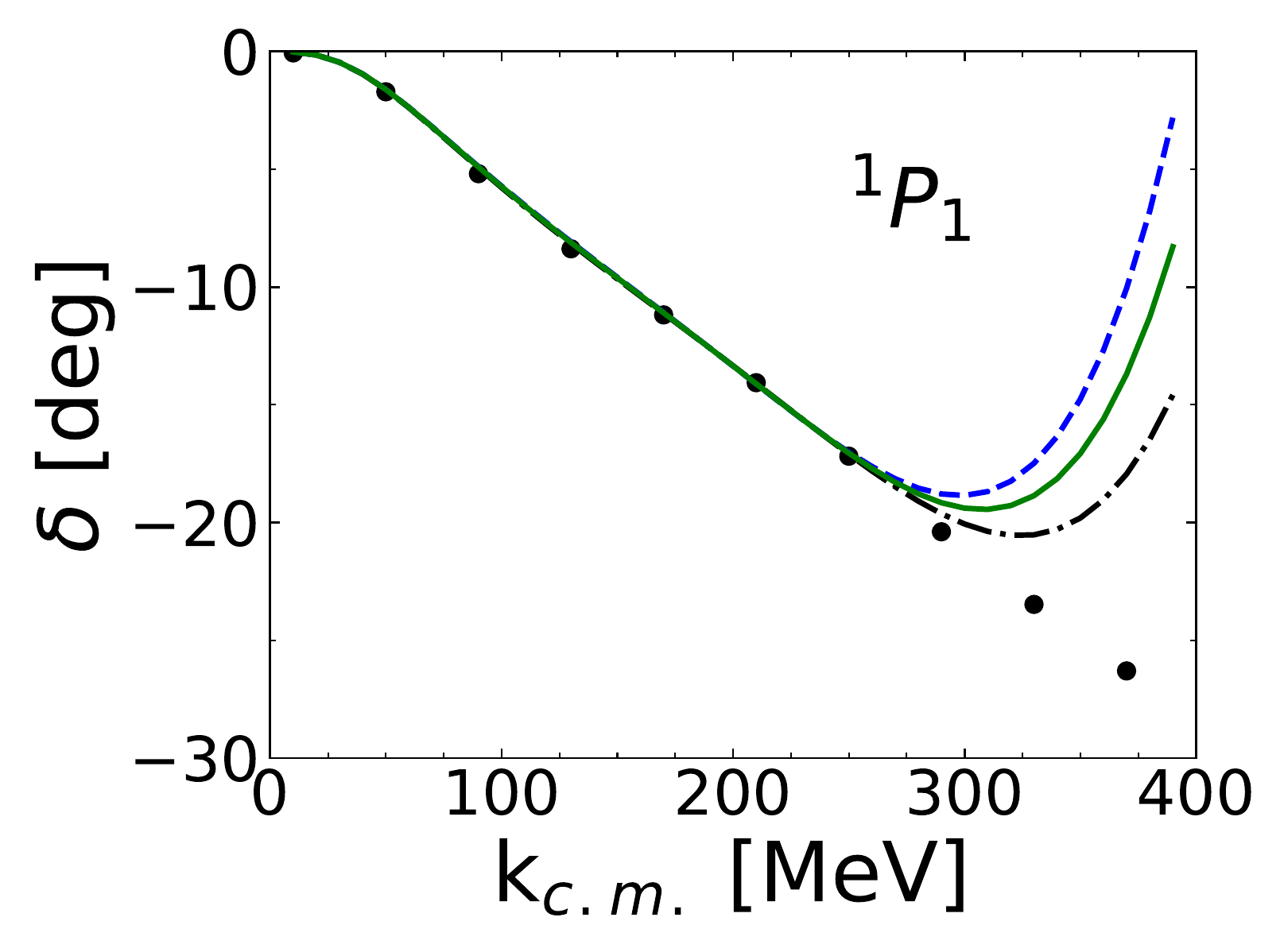}
  \includegraphics[scale=0.30]{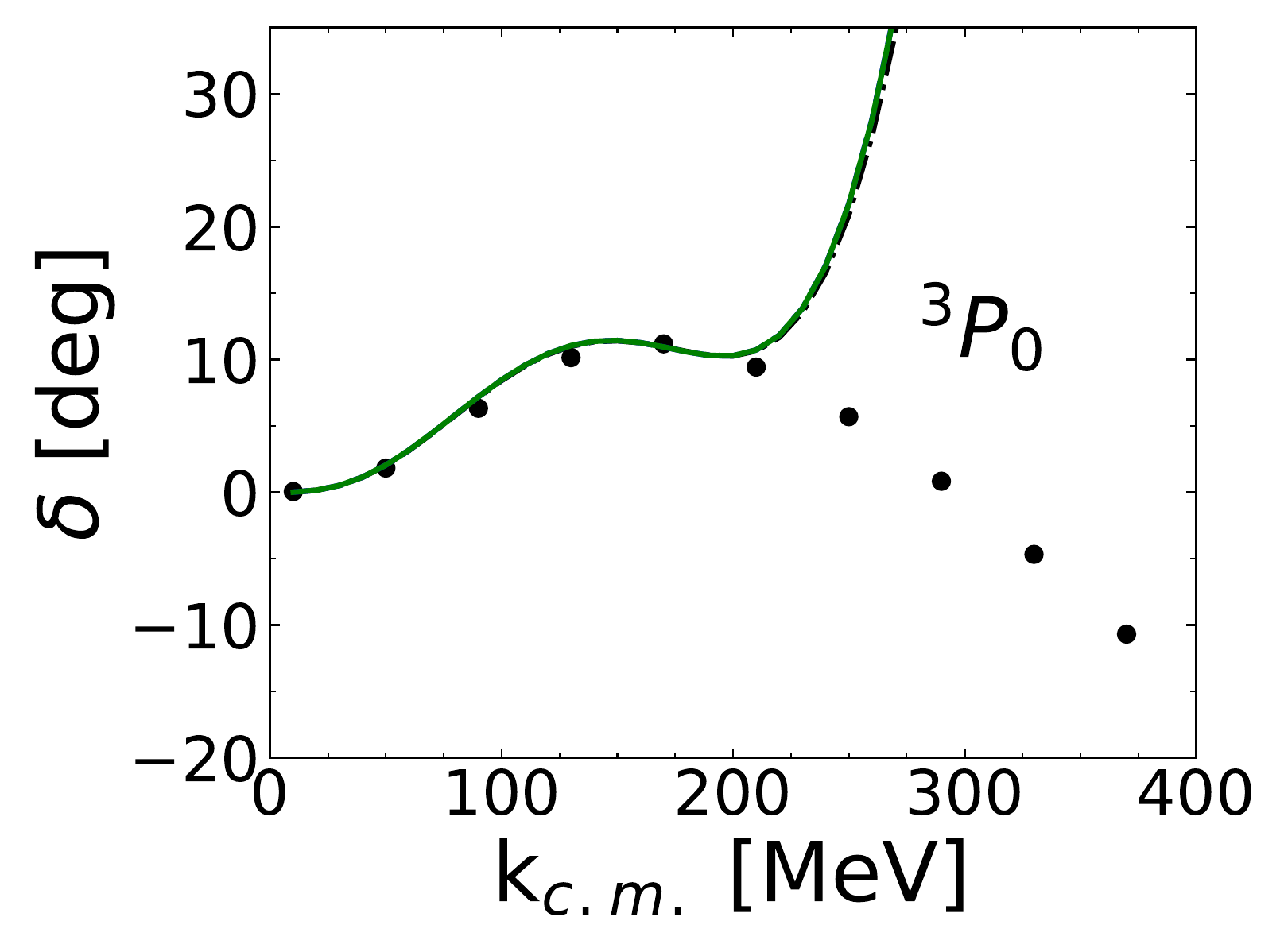}
  \includegraphics[scale=0.30]{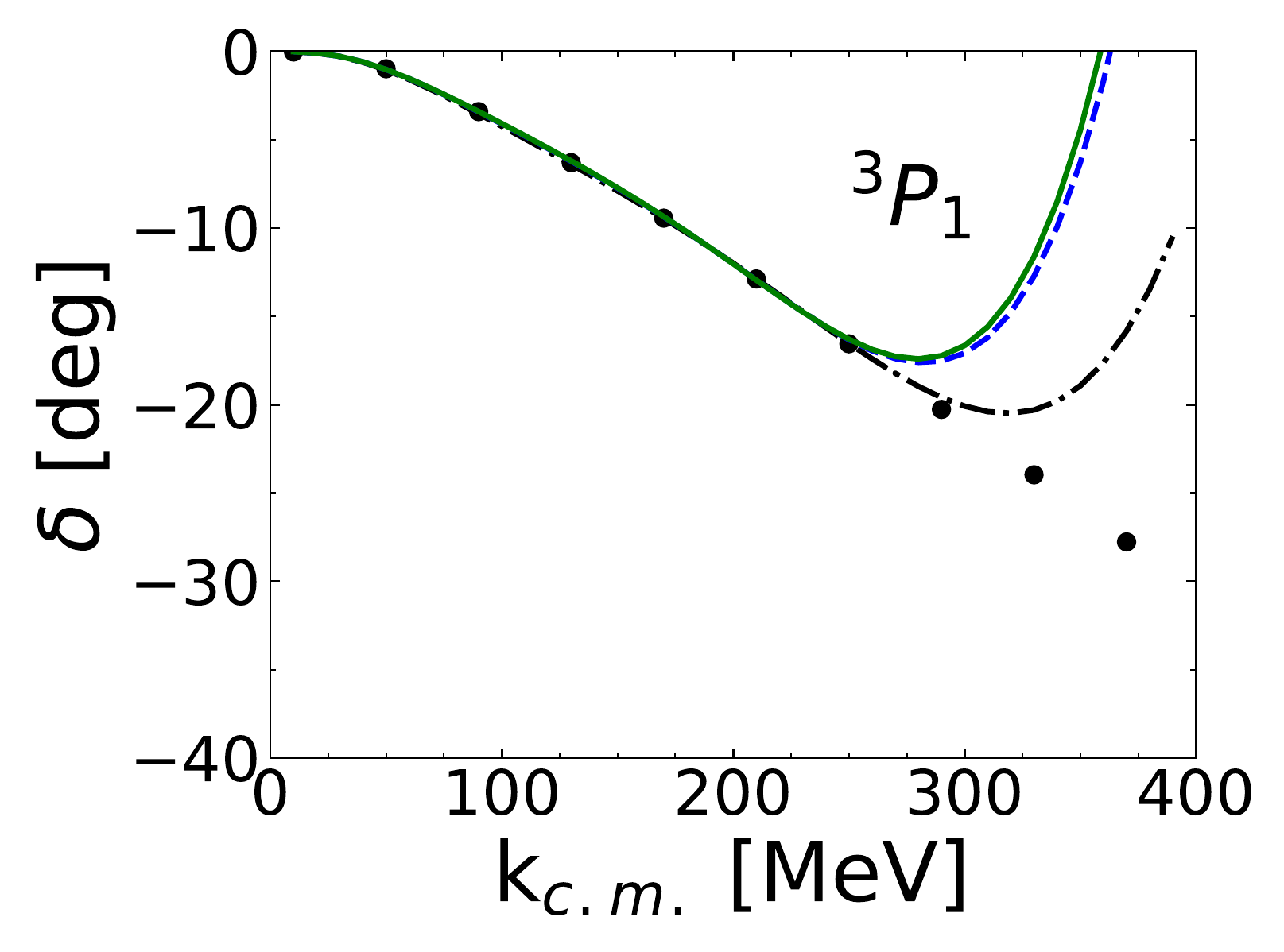}\\
  \includegraphics[scale=0.30]{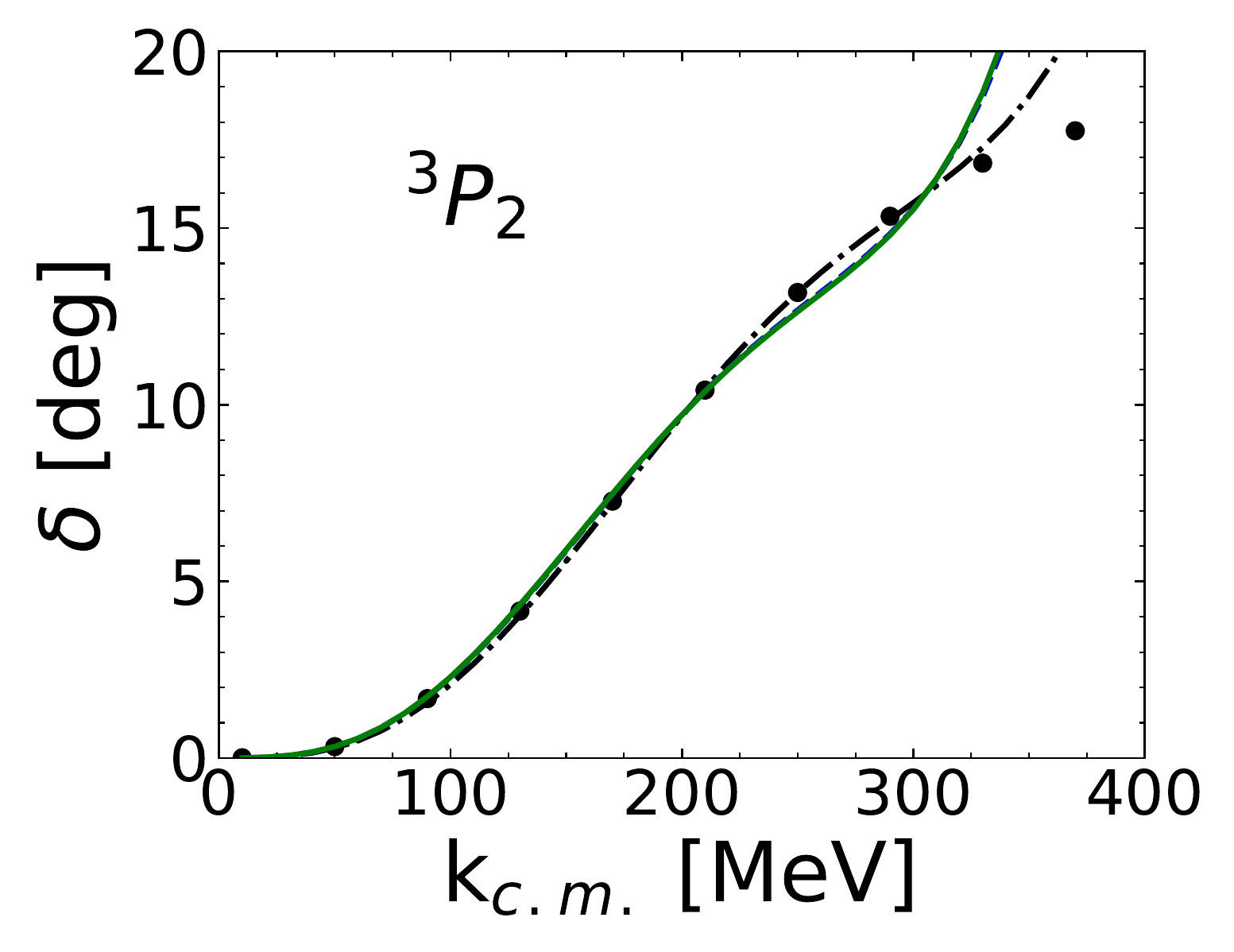}
  \includegraphics[scale=0.30]{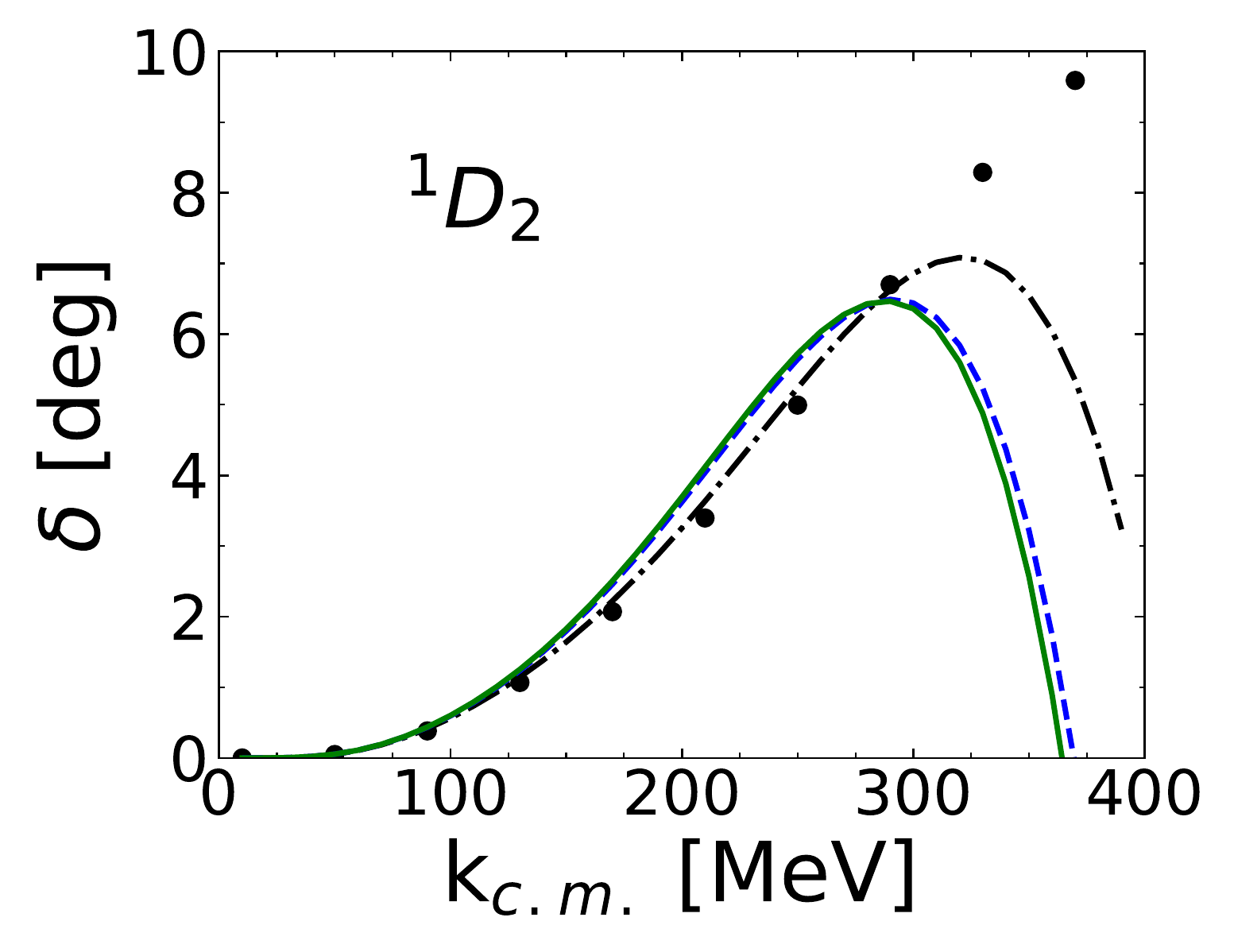}
  \includegraphics[scale=0.30]{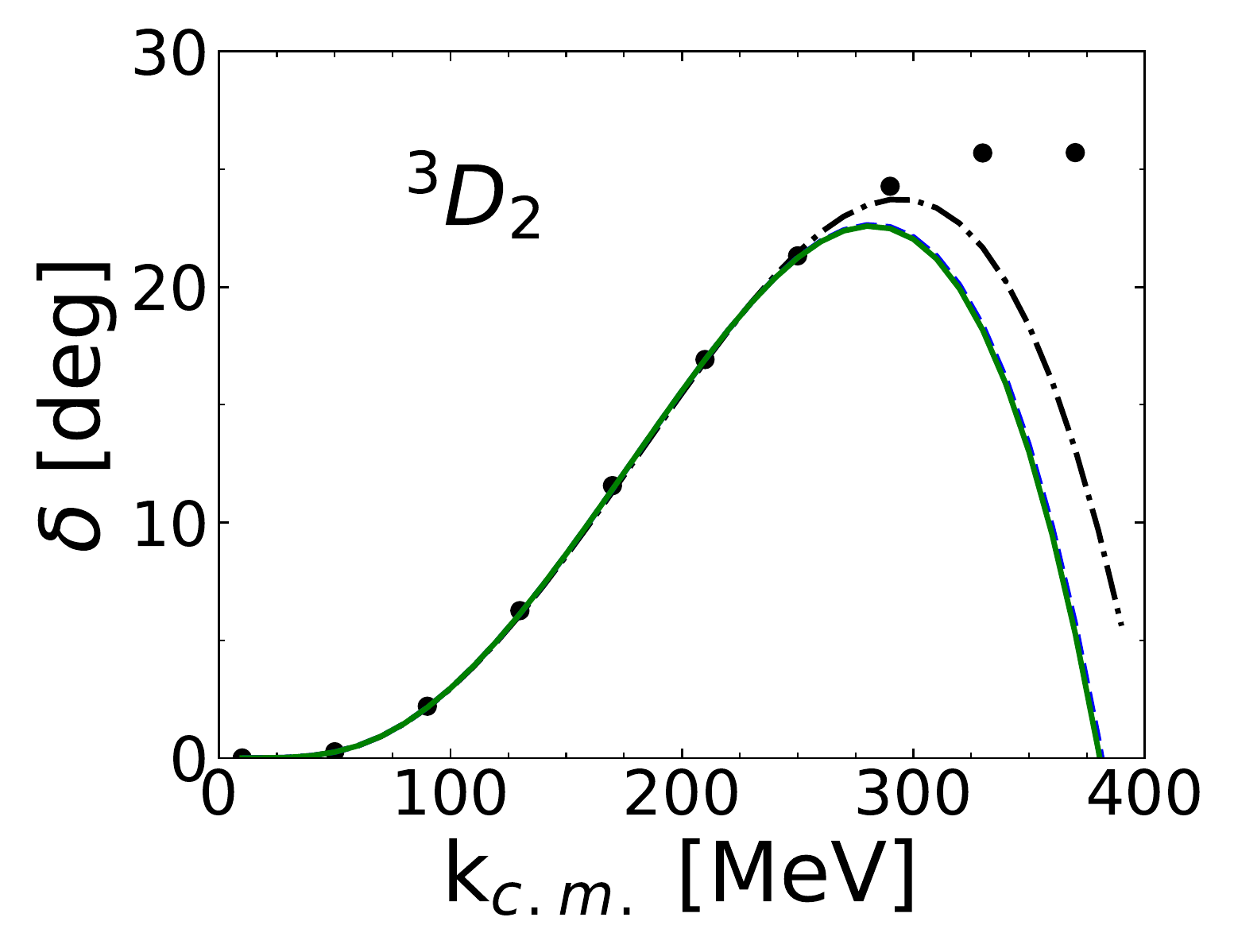}\\
  \includegraphics[scale=0.30]{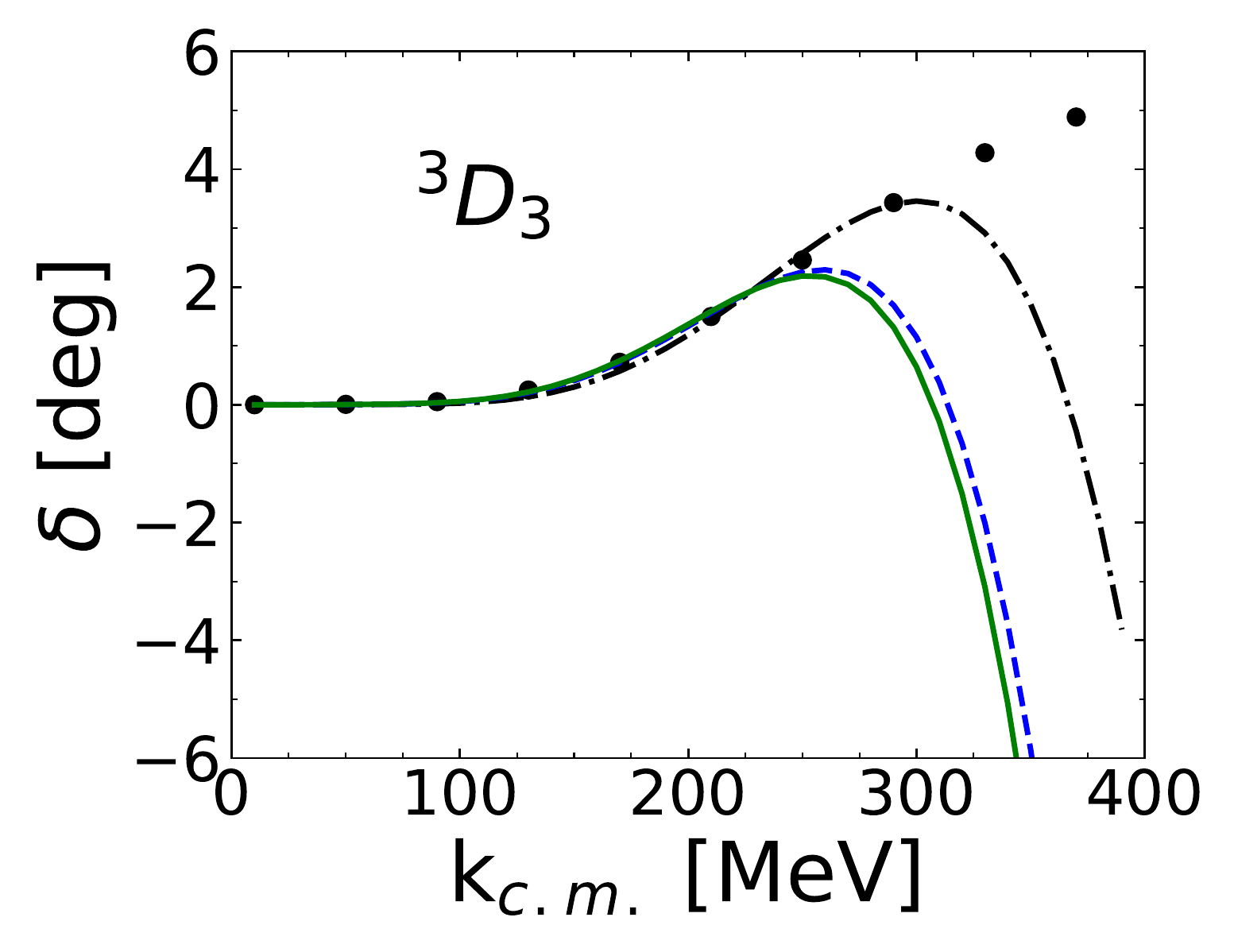}
  \includegraphics[scale=0.30] {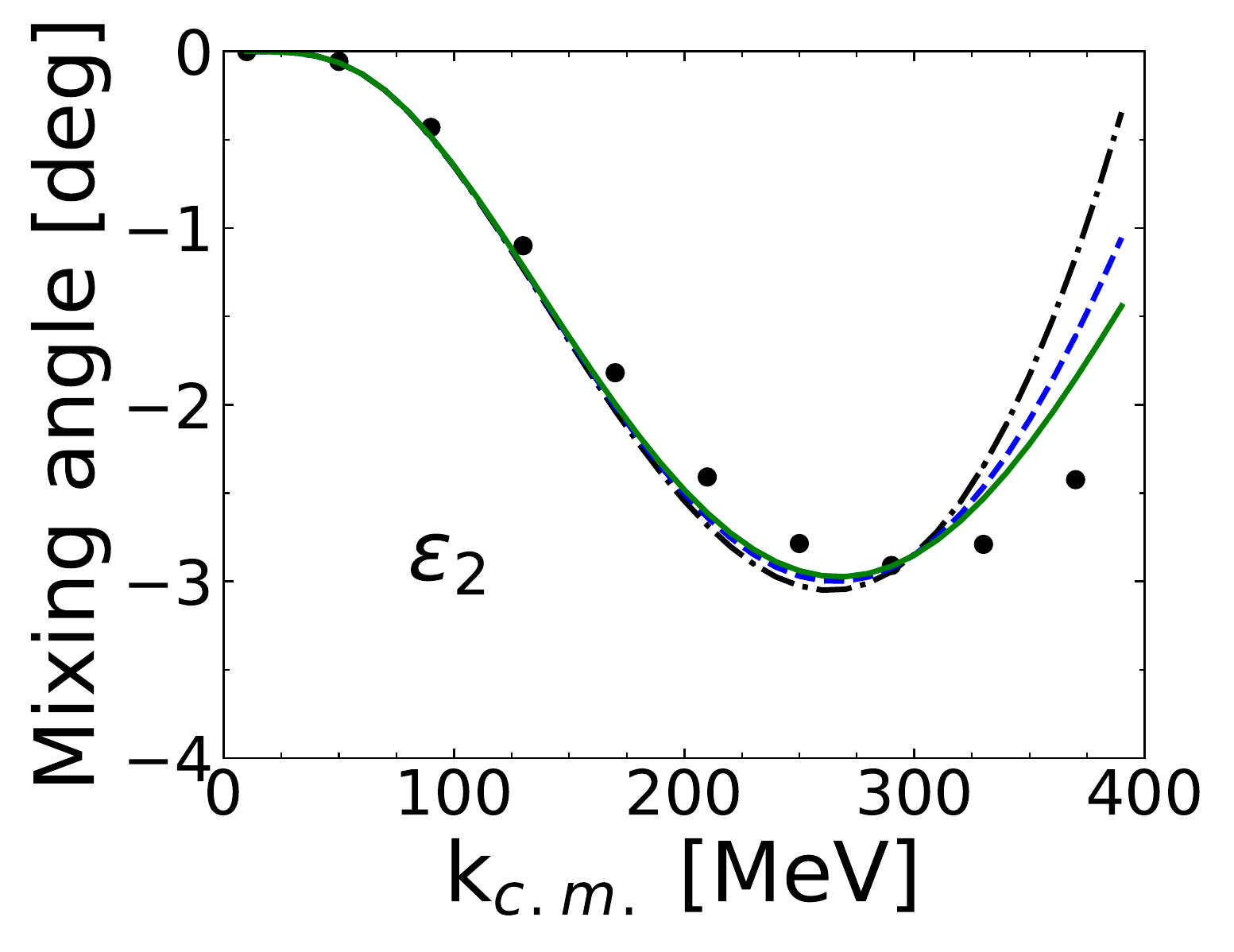}
  \includegraphics[scale=0.30]{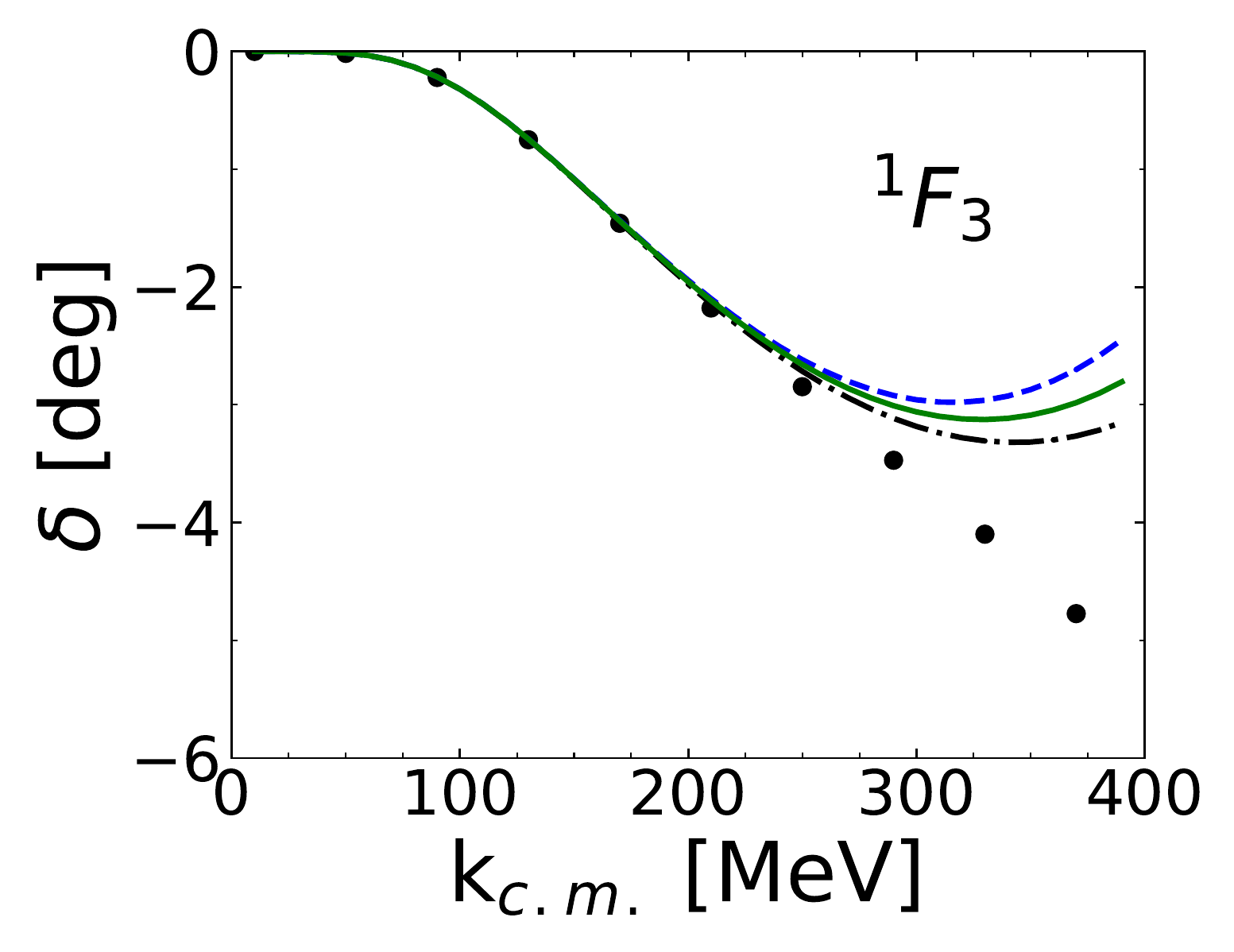}\\
  \includegraphics[scale=0.30]{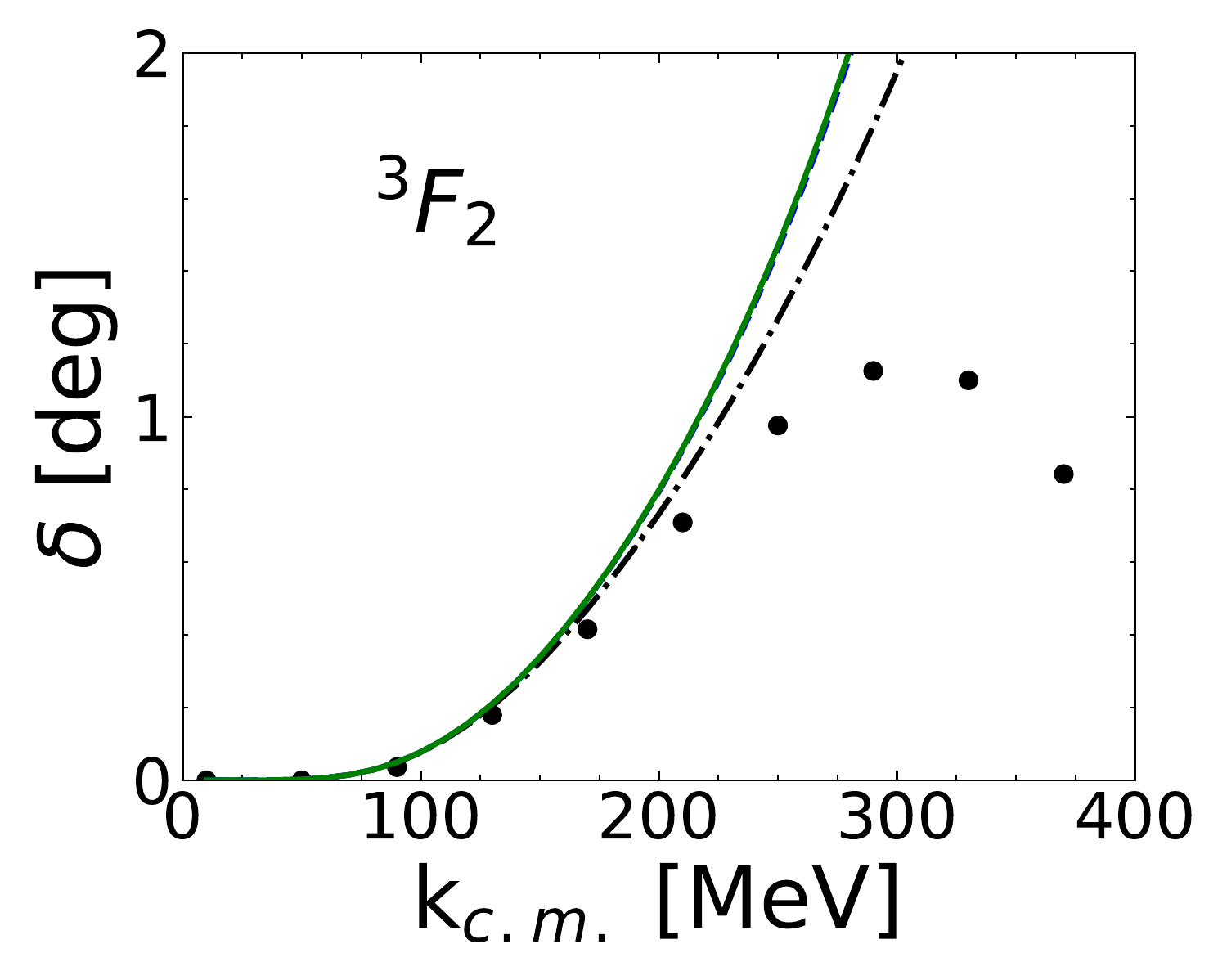}
  \includegraphics[scale=0.30]{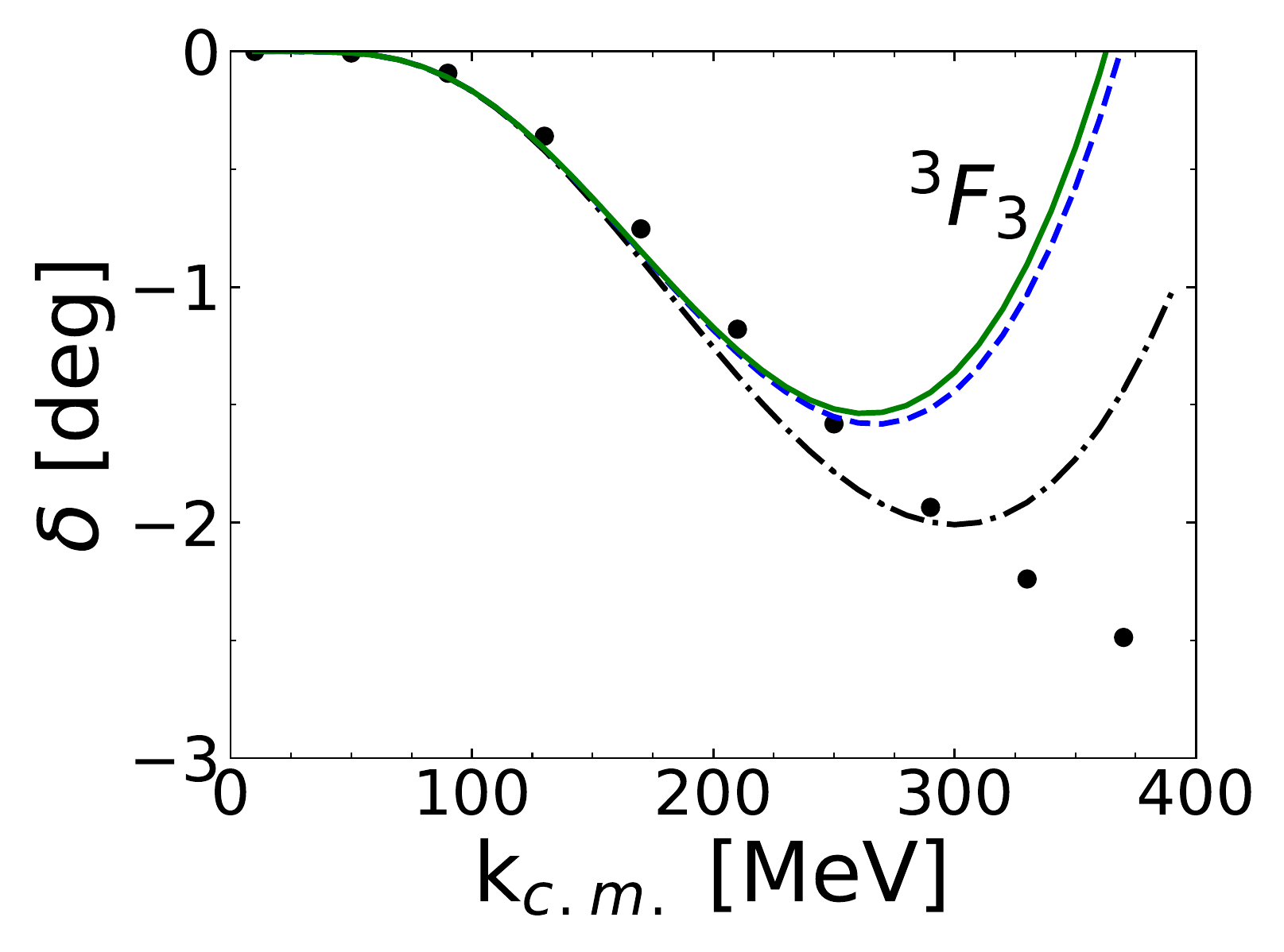}
  \includegraphics[scale=0.30]{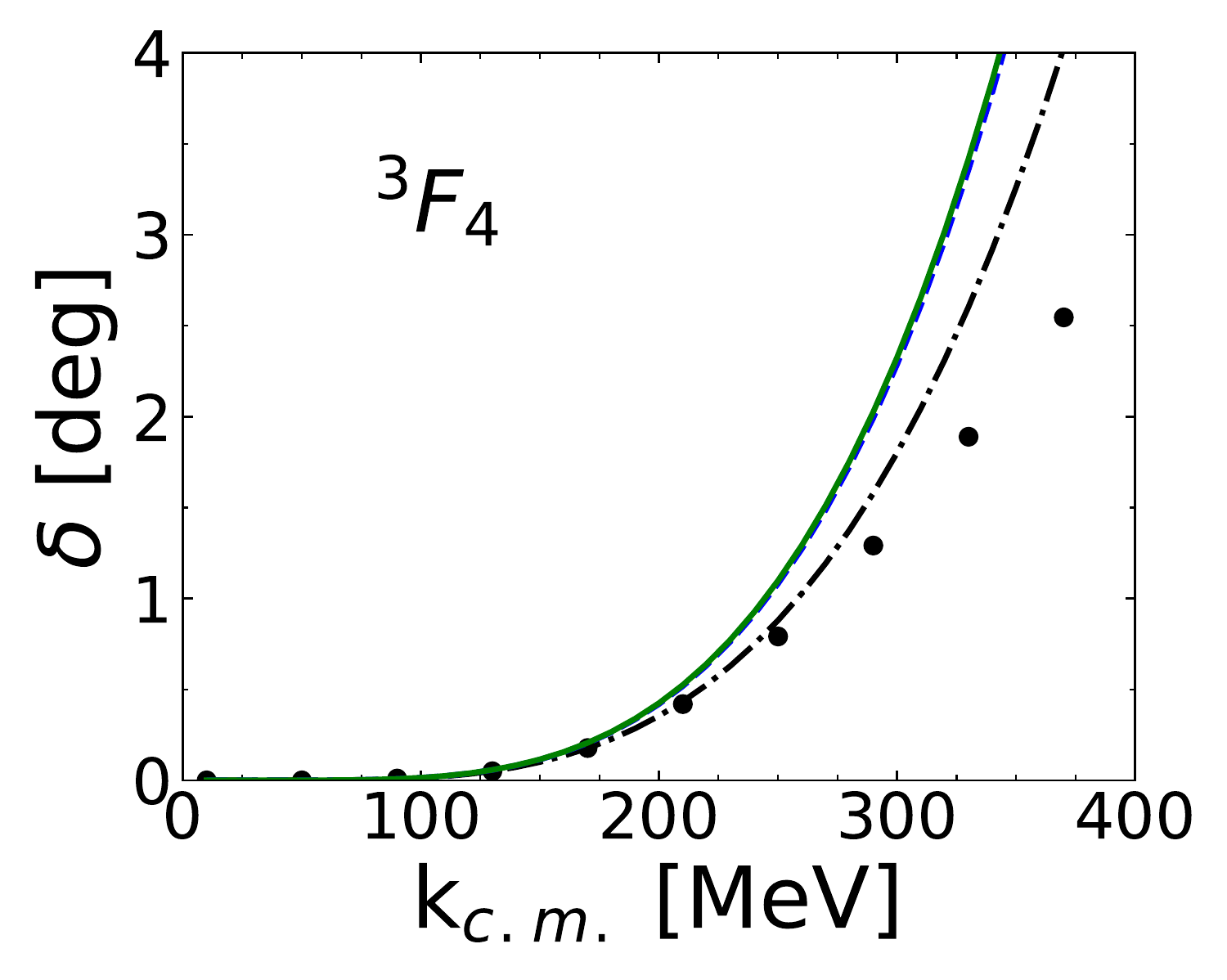}
  \caption{$NN$ phase shifts and mixing angles as functions of the c.m. momentum. The circles are the GWU empirical values, and the lines are the N$^4$LO EFT phase shifts with different sets of $c_i$'s from Table~\ref{tab:cis}: dash-dotted (NLO), dashed (N$^2$LO), and solid (N$^3$LO). $\Lambda = 1$ GeV for all EFT calculations.}
  \label{fig:cisAllWaves}
\end{figure}

Figure~\ref{fig:cisAllWaves} shows how the uncertainty of $c_i$'s affects the N$^4$LO $NN$ phase shifts. A particular value of momentum cutoff is chosen $\Lambda = 1$ GeV, for the cutoff variation is typically much smaller than the uncertainty caused by $c_i$'s. The perturbative OPE expansion still breaks down in $\cp{3}{0}$ at the same $k_\text{c.m.}$ no matter which set of $c_i$ is used, echoing our previous statement that it is the strong attraction of OPE that fails the perturbative scheme rather than anything else. Secondly and probably most importantly, the $c_i$ variation always appear around $k_\text{c.m.} \sim \delta \simeq 300$ MeV, consistent with the anticipation that the delta-less chiral forces step outside their validity range around such momentum scale.

\section{Summary and outlook\label{sec:discussion}}

We have applied perturbative formulation to $NN$ scattering in partial waves with $1 \leqslant L \leqslant 4$, looking into how well perturbative treatment of OPE can describe NN phase shifts. In our notation, LO is reserved for nonperturbative channels, so the nonvanishing amplitude for perturbative channels starts to appear at NLO, the Born approximation of OPE:
\begin{equation}
  T^\text{NLO} = V_{1\pi} \, .
\end{equation}
TPEs are too assumed suppressed by the centrifugal barrier by one order, so that the relative order of $Q^2$ between OPE and TPEs stays unchanged.

The most notable additions compared with previous studies on peripheral waves~\cite{Kaiser-1997mw, Kaiser-1998wa, Entem-2002sf, Birse-2003nz, Epelbaum-2003gr, Krebs-2007rh, Entem-2014msa, Batista-2017vao, RuizSimo-2017anp} include the multiple iterations of OPE (up to three), and iterations involving both OPE and the leading TPE. Contact interactions in $P$ and $D$ waves were also systematically included.

The key takeaway from the paper is that except for $\cs{1}{0}$, $\csd$, and $\cp{3}{0}$, all other channels can be treated in perturbation theory. In particular, our calculation with the delta-less TPEs achieves good agreement with the empirical phase shifts up to $k_\text{c.m.} \simeq 300$ MeV. The uncertainty of $\nu = 1$ $\pi \pi NN$ seagull couplings $c_i$'s from Refs.~\cite{Hoferichter-2015hva, Hoferichter-2015tha} was made use of to examine at what momenta $NN$ data start to sense the delta-isobar. The results gave support to the expected breakdown scale of the delta-less chiral forces $k_\text{c.m.} \sim \delta \simeq 300$ MeV. This suggests a future application of this perturbative scheme on chiral forces with explicit delta-isobar degrees of freedom.

\acknowledgments

B.w.L. thanks Bira van Kolck for useful discussions and the Institut de Physique Nucl\'eaire d'Orsay for hospitality when part of the work was carried out there. The work was supported in part by the National Natural Science Foundation of China (NSFC) under Grant No. 11775148 and No.11735003.

\end{document}